\documentclass[twocolumn]{aastex7}

\usepackage{amsmath}

\shorttitle{\textsc{PhotoIFU}: NIRCam as a Photometric IFU}
\shortauthors{Y.\ Zhu et al.}

\begin{document}

\title{\textsc{PhotoIFU}: NIRCam as a Photometric Integral Field Unit for Mapping Feedback in Galaxies}

\author[0000-0003-3307-7525]{Yongda Zhu} \thanks{JASPER Scholar}
\affiliation{Steward Observatory, University of Arizona, 933 North Cherry Avenue, Tucson, AZ 85721, USA}
\email[show]{yongdaz@arizona.edu}

\author[0000-0002-7893-6170]{Marcia J. Rieke}
\affiliation{Steward Observatory, University of Arizona, 933 North Cherry Avenue, Tucson, AZ 85721, USA}
\email{mrieke@arizona.edu}

\author[0000-0001-6301-3667]{Courtney Carreira}
\affiliation{Department of Astronomy and Astrophysics, University of California, Santa Cruz, 1156 High Street, Santa Cruz, CA 95064, USA}
\email{ccarreir@ucsc.edu}

\author[0000-0001-6561-9443]{Yang Sun}
\affiliation{Steward Observatory, University of Arizona, 933 North Cherry Avenue, Tucson, AZ 85721, USA}
\email{sunyang@arizona.edu}

\author[0000-0001-7673-2257]{Zhiyuan Ji} 
\affiliation{Steward Observatory, University of Arizona, 933 North Cherry Avenue, Tucson, AZ 85721, USA}
\email{zhiyuanji@arizona.edu}

\author[0000-0002-6221-1829]{Jianwei Lyu}
\affiliation{Steward Observatory, University of Arizona, 933 North Cherry Avenue, Tucson, AZ 85721, USA}
\email{jianwei@arizona.edu}

\author[0000-0003-2919-7495]{Christina C.\ Williams}
\affiliation{NSF National Optical-Infrared Astronomy Research Laboratory, 950 North Cherry Avenue, Tucson, AZ 85719, USA}
\email{christina.williams@noirlab.edu}

\author[0000-0002-8909-8782]{Stacey Alberts}
\affiliation{AURA for the European Space Agency (ESA), Space Telescope Science Institute, 3700 San Martin Dr., Baltimore, MD 21218, USA}
\email{salberts@stsci.edu}

\author[0000-0003-3564-6437]{Feng Yuan}
\affiliation{Center for Astronomy and Astrophysics and Department of Physics, Fudan University, Shanghai 200438, China}
\email{fyuan@fudan.edu.cn}

\author[0009-0006-4662-3053]{Yuxuan Zou}
\affiliation{Shanghai Astronomical Observatory, Chinese Academy of Sciences, 80 Nandan Road, Shanghai 200030, China}
\affiliation{University of Chinese Academy of Sciences, No. 19A Yuquan Road, Beijing 100049, China}
\email{yxzou@shao.ac.cn}

\author[0000-0002-0984-7713]{Yurina Nakazato}
\affiliation{Center for Computational Astrophysics, Flatiron Institute, 162 5th Avenue, New York, NY 10010, USA}
\email{ynakazato@flatironinstitute.org}

\author[0000-0002-8651-9879]{Andrew J.\ Bunker}
\affiliation{Department of Physics, University of Oxford, Denys Wilkinson Building, Keble Road, Oxford OX1 3RH, UK}
\email{andy.bunker@physics.ox.ac.uk}

\author[0000-0002-8224-4505]{Sandro Tacchella}
\affiliation{Kavli Institute for Cosmology, University of Cambridge, Madingley Road, Cambridge, CB3 0HA, UK}
\affiliation{Cavendish Laboratory, University of Cambridge, 19 JJ Thomson Avenue, Cambridge, CB3 0HE, UK}
\email{st578@cam.ac.uk}

\author[0000-0002-4622-6617]{Fengwu Sun}
\affiliation{Center for Astrophysics $|$ Harvard \& Smithsonian, 60 Garden St., Cambridge, MA 02138, USA}
\email{fengwu.sun@cfa.harvard.edu}

\author[0000-0003-2303-6519]{George H.\ Rieke}
\affiliation{Steward Observatory, University of Arizona, 933 North Cherry Avenue, Tucson, AZ 85721, USA}
\email{grieke@arizona.edu}

\author[0000-0003-1344-9475]{Eiichi Egami}
\affiliation{Steward Observatory, University of Arizona, 933 North Cherry Avenue, Tucson, AZ 85721, USA}
\email{egami@arizona.edu}

\author[0000-0002-7636-0534]{Jacopo Chevallard}
\affiliation{Department of Physics, University of Oxford, Denys Wilkinson Building, Keble Road, Oxford OX1 3RH, UK}
\email{chevalla@iap.fr}

\author[0000-0003-4565-8239]{Kevin Hainline}
\affiliation{Steward Observatory, University of Arizona, 933 North Cherry Avenue, Tucson, AZ 85721, USA}
\email{kevinhainline@arizona.edu}

\author[0009-0003-5402-4809]{Zheng Ma}
\affiliation{Steward Observatory, University of Arizona, 933 North Cherry Avenue, Tucson, AZ 85721, USA}
\email{mazh@arizona.edu}

\author[0000-0003-4528-5639]{Pablo G. P\'erez-Gonz\'alez}
\affiliation{Centro de Astrobiolog\'ia (CAB), CSIC–INTA, Cra. de Ajalvir Km.~4, 28850- Torrej\'on de Ardoz, Madrid, Spain}
\email{pgperez@cab.inta-csic.es}

\author[0000-0002-5104-8245]{Pierluigi Rinaldi}
\affiliation{Space Telescope Science Institute, 3700 San Martin Drive, Baltimore, Maryland 21218, USA}
\email{prinaldi@stsci.edu}

\author[0000-0001-5171-3930]{Bruno Rodríguez Del Pino}
\affiliation{Centro de Astrobiolog\'ia (CAB), CSIC–INTA, Cra. de Ajalvir Km.~4, 28850- Torrej\'on de Ardoz, Madrid, Spain}
\email{brunorodriguez85@gmail.com}

\author[0009-0000-5273-7870]{Tristen Shields}
\affiliation{Steward Observatory, University of Arizona, 933 North Cherry Avenue, Tucson, AZ 85721, USA}
\email{tdshield@arizona.edu}

\author[0000-0002-9720-3255]{Meredith Stone}
\affiliation{Steward Observatory, University of Arizona, 933 North Cherry Avenue, Tucson, AZ 85721, USA}
\email{meredithstone@arizona.edu}

\author[0000-0001-9262-9997]{Christopher N.\ A.\ Willmer}
\affiliation{Steward Observatory, University of Arizona, 933 North Cherry Avenue, Tucson, AZ 85721, USA}
\email{cnaw@arizona.edu}

\author[0000-0002-8876-5248]{Zihao Wu}
\affiliation{Center for Astrophysics $|$ Harvard \& Smithsonian, 60 Garden St., Cambridge, MA 02138, USA}
\email{zihao.wu@cfa.harvard.edu}

\author[0000-0002-5367-8021]{Minghao Yue}
\affiliation{Steward Observatory, University of Arizona, 933 North Cherry Avenue, Tucson, AZ 85721, USA}
\email{minghao.astro@gmail.com}

\author[0000-0002-1574-2045]{Junyu Zhang}
\affiliation{Steward Observatory, University of Arizona, 933 North Cherry Avenue, Tucson, AZ 85721, USA}
\email{junyuzhang@arizona.edu}

\begin{abstract}
We present \textsc{PhotoIFU}, a workflow that uses deep multi-band imaging as a low-resolution photometric integral field unit. Applied to PSF-matched JWST/NIRCam imaging, \textsc{PhotoIFU} treats each spatial pixel as a coarse SED element and fits the pixel SEDs with \textsc{Prospector} to map resolved stellar-population and ISM-related properties. We apply this approach to three galaxies at $z=1.3$--3.7 in JADES: two systems with extended ionized line emission and one post-starburst galaxy with an exceptionally strong neutral outflow. Pixel-by-pixel SED fitting gives maps of stellar-mass surface density, specific star formation rate, dust attenuation, gas-phase metallicity, and recent star-formation history. We find that regions selected from the extended-emission or outflow geometry occupy distinct parts of the resolved SED-property distribution compared with the full host. In the systems with extended ionized emission, these regions are generally less dusty, consistent with ionized emission being observed along dust-poor, low-column-density pathways through the host. In the neutral-outflow system, the selected regions show enhanced recent star formation, suggesting that compact rejuvenation may mark the aftermath of an earlier energetic phase. These results show that galactic outflows and extended emission-line structures can be spatially associated with measurable differences in resolved host-galaxy stellar populations and ISM-related properties. \textsc{PhotoIFU} provides an imaging-based method for resolved SED mapping of feedback-related structures in larger galaxy samples where full spectroscopic integral-field mapping is unavailable.
\end{abstract}


\section{Introduction}\label{sec:intro}

Galactic feedback is a central part of the baryon cycle. Energy, momentum, metals, and dust released by massive stars and accreting black holes can reshape the interstellar medium (ISM), drive gas into the circumgalactic medium (CGM), and regulate the growth of galaxies \citep[e.g.,][]{veilleux_galactic_2005,somerville_physical_2015,naab_theoretical_2017,veilleux_cool_2020,liu_galactic_2022,zhu_early_2026}. In modern galaxy-formation models, feedback is needed to reproduce basic galaxy properties, including stellar masses, baryon fractions, metal enrichment, and the buildup of quenched systems \citep[e.g.,][]{white_core_1978,crain_eagle_2015,pillepich_first_2018,nelson_first_2019}. Observationally, the key question is no longer simply whether outflows exist. Outflows are now detected across many gas phases and redshifts \citep[e.g.,][]{bischetti_multiphase_2024,liu_frequent_2025,davies_jwst_2024,zhu_potential_2026,sun_extreme_2025,sun_census_2026,zhu_there_2026}. The next question is where the gas signatures of feedback lie within galaxies, and whether those regions differ from the rest of the host in their stellar-population and ISM-related properties.

Observations have identified and characterized galactic outflows through spectroscopy and line-sensitive imaging, including broad emission lines, blueshifted interstellar absorption, molecular and atomic gas kinematics, and extended ionized emission \citep[e.g.,][]{weiner_ubiquitous_2009,forster_schreiber_kmos3d_2019,ginolfi_alpine-alma_2020,butler_molecular_2023,perrotta_kinematics_2023,xu_stellar_2023,zhu_systematic_2025}. These studies show that outflows occur across multiple gas phases and are common in diverse galaxy and AGN populations \citep[e.g.,][]{zhu_early_2026,zhu_potential_2026,zhu_smiles_2026,sun_census_2026}. Spatially resolved observations also show that the gas and radiation fields are frequently anisotropic or bipolar \citep[e.g.,][]{guo_bipolar_2023,lebowitz_jwst_2025,lebowitz_spatially_2026}. Spatially resolved spectroscopic studies show that outflows have kiloparsec-scale structure and can be associated with galactic nuclei, disks, and star-forming regions \citep[e.g.,][]{venturi_magnum_2021,ren_rioja_2025,venturi_ga-nifs_2025,ulivi_ga-nifs_2026,rodriguez_del_pino_ga-nifs_2026}. Cold molecular outflows provide a complementary view of feedback in another gas phase \citep{fluetsch_cold_2019}. Besides IFU, deep JWST slitless spectroscopy and medium-band imaging have begun to reveal extended ionized structures directly, including lensed [O\,\textsc{iii}] filaments in the CGM of a massive system at $z\sim2.8$ \citep{peng_direct_2025}. Theoretical work likewise shows that feedback propagates through an inhomogeneous ISM and produces spatially and temporally variable gas flows \citep[e.g.,][]{hopkins_galaxies_2014,muratov_gusty_2015,nelson_first_2019}. The gas and dust geometry regulates the directions through which ionizing radiation can escape, with dust-poor, low-column-density channels dominating the emergent radiation field in some systems \citep{ji_importance_2025,menon_bursts_2025}. Recent high-resolution simulations further suggest that feedback, bursty star formation, gas compaction, and short quenching or rejuvenation episodes can be closely coupled during early galaxy growth \citep[e.g.,][]{ceverino_effects_2023,dome_mini-quenching_2024,mcclymont_thesan-zoom_2025}.

A major observational gap remains. Spectroscopy gives the cleanest view of gas kinematics, line ratios, and absorption, but spatially complete spectroscopy is difficult to obtain for large samples at high redshift. JWST/NIRSpec IFU observations are powerful but currently limited to small samples, as shown by recent GA-NIFS and RIOJA studies of high-redshift galaxies and AGN \citep[e.g.,][]{ubler_ga-nifs_2023,deugenio_fast-rotator_2024,lamperti_ga-nifs_2024,bertola_ga-nifs_2025,mawatari_rioja_2025,ren_rioja_2025,venturi_ga-nifs_2025,rodriguez_del_pino_ga-nifs_2026}. NIRSpec MSA surveys reach many galaxies, but the spectra usually cover only the part of each galaxy within the designed slitlets; slit-stepping programs provide an efficient intermediate approach for spatially resolved spectroscopy \citep[see also][]{morrison_creating_2025,ju_msa-3d_2025}. Imaging, on the other hand, provides complete spatial coverage, but broad-band morphology alone cannot determine whether a color-selected structure is also distinct in stellar mass, recent star formation, dust attenuation, or metallicity. A practical path is to combine gas signatures identified from spectroscopy or line-sensitive imaging with deep multi-band NIRCam photometry that maps the resolved SED properties of the same system.

With JWST/NIRCam \citep{gardner_james_2023,rieke_performance_2023}, the wide filters provide low-resolution spectral sampling with effective $R\sim4$--5, while the medium filters reach $R\sim10$--20. With the depth, wavelength coverage, and angular resolution of the JWST Advanced Deep Extragalactic Survey (JADES; \citealp{rieke_jades_2023,eisenstein_overview_2023,bunker_jades_2024,eisenstein_jades_2025,robertson_jwst_2026}), these data can be used as a photometric integral field unit: each spatial pixel has a coarse SED that can be fit for physical properties. We refer to this workflow as \textsc{PhotoIFU}: using PSF-matched multi-band imaging as a low-resolution photometric integral field unit. We note that spatially resolved SED fitting itself is an established approach. General resolved-source tools such as \textsc{piXedfit} provide infrastructure for fitting resolved photometry \citep{abdurrouf_introducing_2021}. Recent JWST studies have also performed pixel-level or substructure-level SED analyses of high-redshift galaxies \citep[e.g.,][]{perez-gonzalez_ceers_2023,zhu_clumps_2026}. The distinction here is not the use of pixel-by-pixel SED fitting itself, but the combination of gas-signature-guided apertures with a \textsc{Prospector}-like model applied to thousands of pixels. In the implementation used here, the pixel SEDs are fit with non-parametric SFHs, nebular emission, dust attenuation, and gas-phase metallicity, accelerated with the \textsc{Parrot} emulator. This setup is useful because the quantities of interest include recent SFH, line-sensitive nebular properties, and dust attenuation. The medium-band search of \citet{zhu_systematic_2025} showed that NIRCam can identify extended line-emission structures by comparing line-sensitive filters with adjacent continuum filters. Here we use selected systems from the same imaging data to ask whether regions associated with ionized or neutral gas signatures occupy distinct parts of the resolved SED-property distribution.

This paper is a pilot study of three representative systems with gas signatures identified from spectroscopy and/or medium-band line-emission morphology. We use PSF-matched JADES NIRCam imaging and pixel-by-pixel SED fitting to map stellar mass, specific star formation rate, dust attenuation, gas-phase metallicity, and recent-to-past star-formation history. We then compare regions selected from extended emission or outflow geometry with the full host distribution. Section \ref{sec:data} describes the sample and NIRCam data. Section \ref{sec:methods} presents the image preparation and spatially resolved SED fitting. Section \ref{sec:results} gives the resolved-map and region-comparison results. Section \ref{sec:discussion} compares the results with simulations, discusses the physical interpretation and limitations, and Section \ref{sec:summary} summarizes the main conclusions. The \textsc{PhotoIFU} workflow will be made available at \url{https://github.com/ydzhuastro/photoifu}.

\section{Sample and Data}\label{sec:data}

\begin{figure*}[!ht]
\centering
\includegraphics[width=\textwidth]{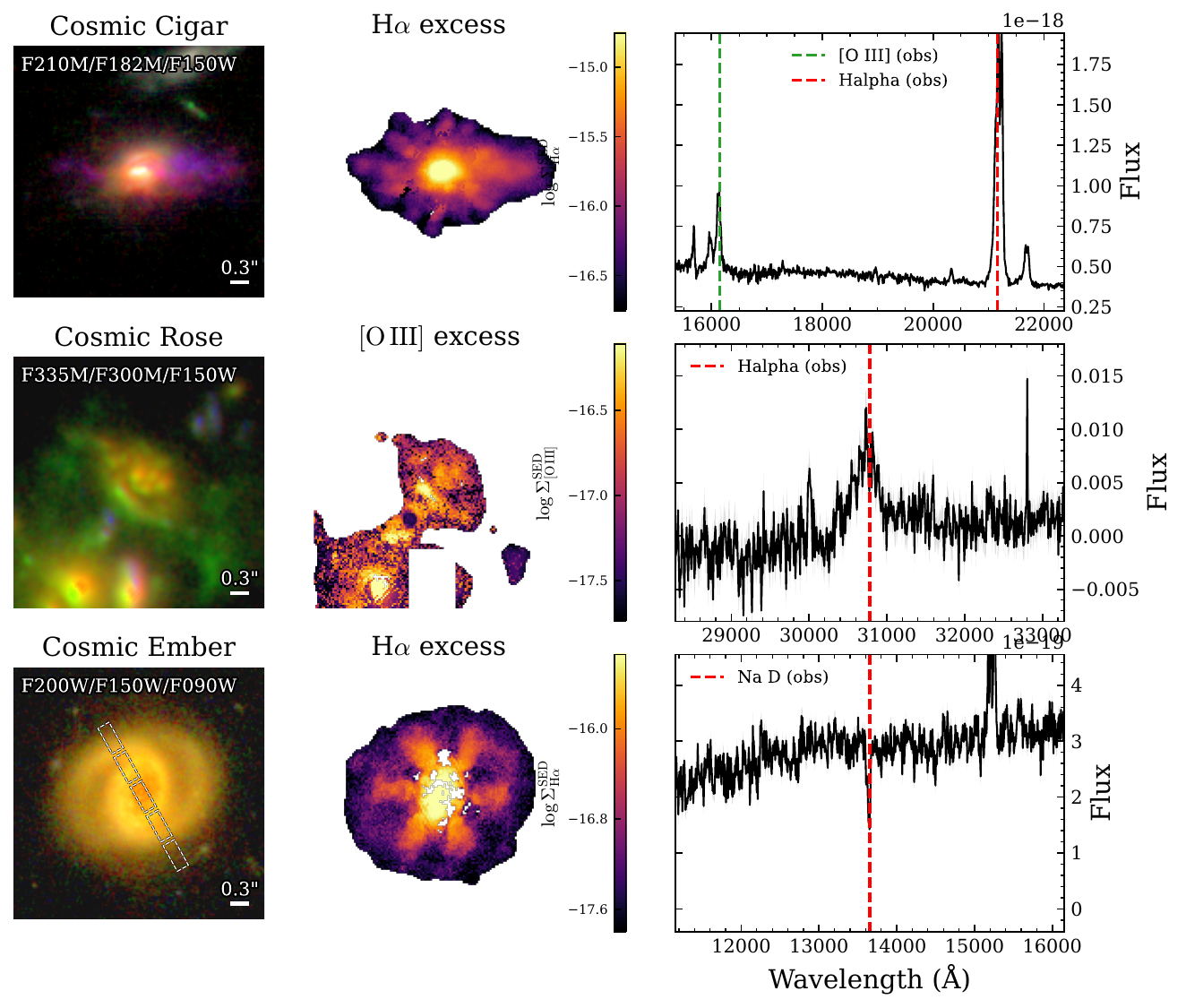}
\caption{False-color images, best-fit line-excess maps, and spectra for the three systems studied in this paper. We show the Cosmic Cigar (JADES 209962, $z=2.2251$), a broad-line AGN host with biconical H$\alpha$ emission; the Cosmic Rose (JADES 172813, $z=3.690$), a clumpy system with remarkably extended [O\,\textsc{iii}] emission; and the Cosmic Ember (JADES/SMILES 206183, $z=1.3171$), a quiescent or post-starburst galaxy with a strong Na\,\textsc{D} neutral outflow. The middle column shows $\Sigma_{\rm line}^{\rm SED}$ maps for H$\alpha$, [O\,\textsc{iii}], and H$\alpha$, respectively. Here $\Sigma_{\rm line}^{\rm SED}$ is the best-fit SED line-excess surface brightness, measured by integrating the continuum-subtracted best-fit SED across the line window and dividing by the $0.03''\times0.03''$ pixel area. The maps are shown in units of ${\rm erg~s^{-1}~cm^{-2}~arcsec^{-2}}$. White dashed rectangles in the Cosmic Ember RGB panel mark the approximate NIRSpec MSA shutter layout for the SMILES spectrum. The spectra come from the SMILES NIRSpec MSA data for the Cosmic Cigar and the Cosmic Ember and from NIRCam wide-field slitless grism data for the Cosmic Rose, as described in the text.}
\label{fig:rgb_spectra}
\end{figure*}

We focus on three galaxies with gas signatures identified from spectroscopy and/or medium-band line-emission morphology. The sample represents three complementary cases: an ionized outflow around a broad-line AGN, extended ionized emission in a clumpy system at $z=3.690$ without an obvious AGN in the central component, and a powerful neutral outflow in a post-starburst galaxy with no clear evidence for ongoing luminous AGN activity. Together, these systems allow us to test whether gas-selected structures are spatially connected to distinct stellar-population and ISM-related properties within those systems.

\begin{itemize}
    \item JADES-GS-209962, the ``Cosmic Cigar'', lies at $z=2.2251$ and is a broad-line and MIRI-selected AGN host with a biconical H$\alpha$/ionized-emission structure. It was identified by the medium-band candidate search as a strong example of extended H$\alpha$ emission \citep{zhu_systematic_2025} and is discussed as part of the MIRI AGN sample in \citet{rieke_confirming_2025}. It is also known as K20-ID5 \citep{forster_schreiber_sinszc-sinf_2014,genzel_evidence_2014,loiacono_multiwavelength_2019,scholtz_kashz_2020,davies_nuclear_2020}.

    \item JADES-GS-172813, the ``Cosmic Rose'' \citep{williams_galaxies_2024,alberts_high_2024,zhu_systematic_2025}, lies at $z=3.690$ and shows a clumpy, multi-component morphology together with remarkably extended [O\,\textsc{iii}] emission. The central system does not show evidence for an AGN, while the lower-left, southeastern compact component is likely to host an AGN given the Chandra and MIRI detection (J.\ Lyu et al. in prep). The Cosmic Rose provides a useful case for testing whether large-scale ionized emission is connected to spatially distinct stellar-population and ISM-related properties across a clumpy galaxy system.

    \item JADES-GS-206183, the ``Cosmic Ember'', lies at $z=1.3171$ and is a quiescent or post-starburst galaxy with strong Na\,\textsc{D} absorption indicating an extreme neutral outflow with $v_{\rm out}\sim828~{\rm km~s^{-1}}$ and $\log \dot{M}_{\rm out}\sim2.40~M_\odot~{\rm yr^{-1}}$ \citep{sun_extreme_2025}. The galaxy shows no clear evidence for ongoing luminous AGN activity, and its present global SFR is below that required to explain the inferred outflow in the integrated analysis of \citet{sun_extreme_2025}. It therefore provides a fossil-outflow case, where the gas signature likely records an earlier, more energetic phase of galaxy evolution.
\end{itemize}

For reference, integrated SED measurements place all three systems at high stellar mass. The Cosmic Cigar has $\log(M_\star/M_\odot)=10.8$ and ${\rm SFR}_{\rm SED}=225.4~M_\odot~{\rm yr}^{-1}$ in the MIRI-selected AGN analysis of \citet{rieke_confirming_2025}. For the Cosmic Rose, JADES 172813 is a dusty star-forming galaxy with $\log(M_\star/M_\odot)\simeq11$, ${\rm SFR}\sim730~M_\odot~{\rm yr}^{-1}$, and $A_V\sim3.5$ \citep{alberts_high_2024}. The Cosmic Ember has $\log(M_\star/M_\odot)=11.15^{+0.04}_{-0.05}$ and ${\rm SFR}_{\rm SED}=7.5^{+4.1}_{-2.1}~M_\odot~{\rm yr}^{-1}$ \citep{sun_extreme_2025}. Summing the fitted \textsc{PhotoIFU} pixels gives $\log(M_\star/M_\odot)=10.82$, 10.96, and 10.93 for the Cosmic Cigar, the Cosmic Rose, and the Cosmic Ember, respectively, with corresponding 0--30 Myr SFRs of 210, 340, and $34~M_\odot~{\rm yr}^{-1}$. These summed pixel values use the 0--30 Myr SFR bin and are included only to summarize the fitted maps; the analysis below is based on differential comparisons among pixels within each target.

Figure~\ref{fig:rgb_spectra} shows the imaging and spectroscopic observations for the three systems. We use JADES DR5 NIRCam imaging \citep{johnson_jwst_2026,robertson_jwst_2026,carreira_jwst_2026}, together with spectroscopy that anchors the systemic redshifts and identifies the outflow signatures in each source. For the Cosmic Cigar, NIRSpec F100LP/G140M and F170LP/G235M spectroscopy from SMILES (PID~1207, PI: G.~Rieke; \citealp{rieke_smiles_2024,alberts_smiles_2024,zhu_smiles_2026}) provides the systemic redshift and shows broad H$\alpha$ and [O\,\textsc{iii}] emission associated with the AGN and ionized outflow. The source was previously shown to host a powerful nuclear ionized outflow in SINFONI spectroscopy \citep{forster_schreiber_sinszc-sinf_2014,genzel_evidence_2014,davies_nuclear_2020}. Its extended H$\alpha$ morphology is identified in the NIRCam medium-band analysis of \citet{zhu_systematic_2025}. For the Cosmic Rose, NIRCam/F322W2 wide-field slitless grism spectroscopy from the JADES Origins Field program (JOF; PID~4540, PI: D.~Eisenstein; \citealp{sun_jades_2026}) provides the redshift and reveals a blueshifted H$\alpha$ component. Together with the extended [O\,\textsc{iii}] morphology, this supports the interpretation of a strong ionized outflow in the system. For the Cosmic Ember, SMILES spectroscopy provides the redshift and reveals blueshifted Na\,\textsc{D} absorption, identifying a powerful neutral outflow in a post-starburst system with no clear evidence for ongoing luminous AGN activity \citep{sun_extreme_2025,zhu_smiles_2026}.

The pixel-level analysis uses PSF-matched NIRCam images registered to a common WCS and convolved to the F444W PSF \citep[see][for details]{johnson_jwst_2026}. The RGB filters shown in Figure~\ref{fig:rgb_spectra} are F210M/F182M/F150W for the Cosmic Cigar, F335M/F300M/F150W for the Cosmic Rose, and F200W/F150W/F090W for the Cosmic Ember. The spatially resolved SED fitting uses the available PSF-matched NIRCam imaging in F090W, F115W, F150W, F182M, F200W, F210M, F250M, F277W, F300M, F335M, F356W, F410M, and F444W. The cutout catalogs contain only pixels passing the host-pixel selection described below, with per-pixel fluxes and uncertainties in each available filter.

\section{SED Fitting}\label{sec:methods}

\begin{figure*}[!ht]
\centering
\includegraphics[width=0.95\textwidth]{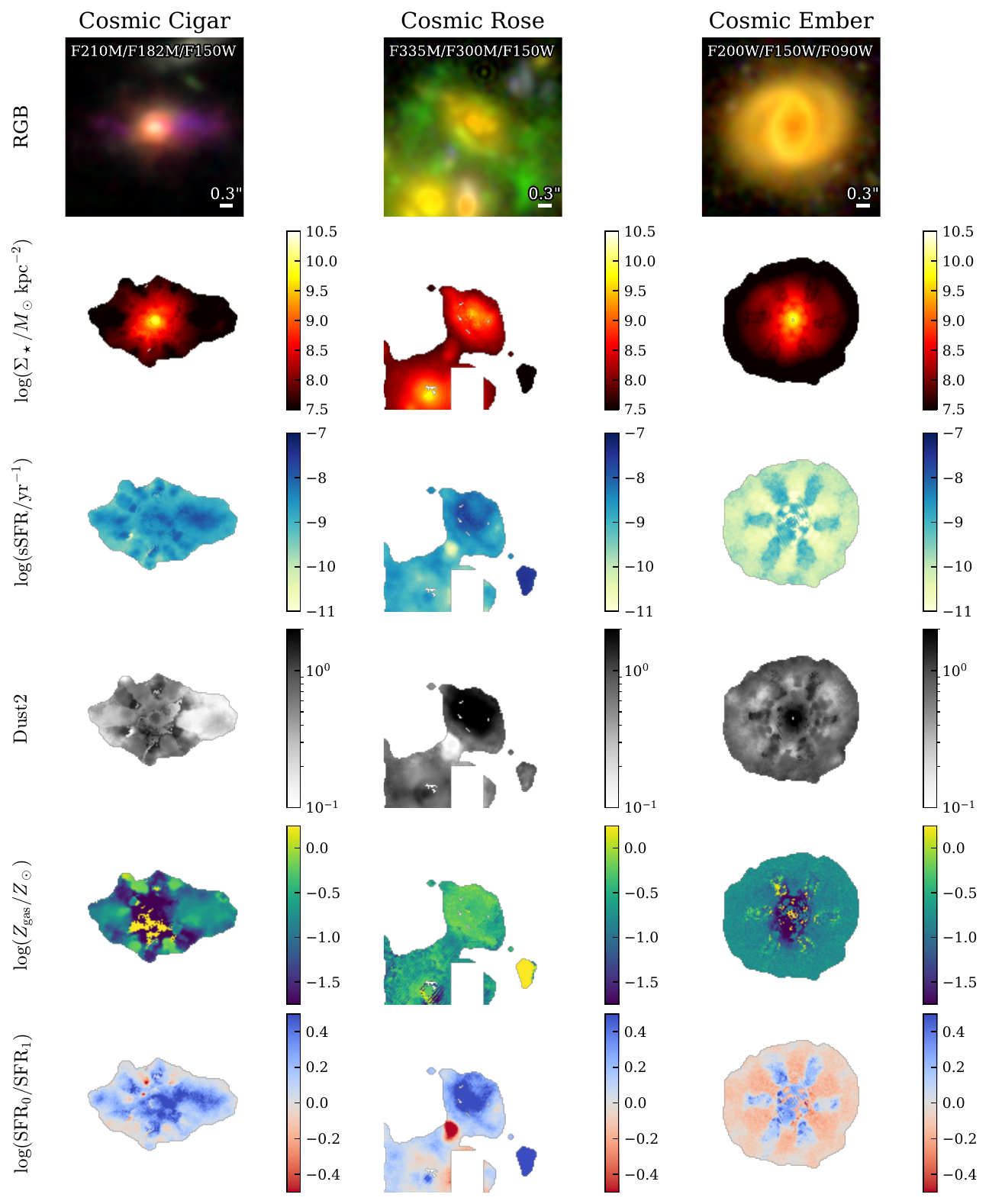}
\caption{Resolved NIRCam SED maps for the three targets. Each column shows one source. Rows show the PSF-matched RGB image and resolved SED-property maps: projected stellar-mass surface density $\log(\Sigma_\star/M_\odot\,{\rm kpc}^{-2})$, specific star formation rate $\log({\rm sSFR}/{\rm yr}^{-1})$, the dimensionless diffuse dust optical-depth normalization \texttt{Dust2}, gas-phase metallicity $\log(Z_{\rm gas}/Z_\odot)$, and the recent-to-past star-formation ratio $\log({\rm SFR}_0/{\rm SFR}_1)$. The per-pixel stellar mass and SFR values are converted to projected physical surface densities using the $0.03''$ pixel scale and the angular-diameter distance of each source. Only valid fitted pixels are shown. Isolated non-finite pixels surrounded by finite neighbors are filled only for display and are not used in the quantitative analysis. The weak radial pattern visible in the Cosmic Ember reflects the F444W-matched PSF and the strong central light concentration, so the spatial extent of compact gradients should be interpreted as an upper limit.}
\label{fig:sed_maps}
\end{figure*}

\subsection{Input Pixel Selection}\label{subsec:imageprep}

For each source, we determine the source position and catalog size information from the JADES photometric catalog, and then extract a square cutout in each available filter. The default cutout is a $132\times132$ pixel image stamp with a pixel size of 0.03\arcsec.

The pixels used in the SED fits are selected from these full image stamps. We first build a detection image for each source. The detection filters are F210M for the Cosmic Cigar, F300M for the Cosmic Rose, and F150W for the Cosmic Ember. F210M and F300M are line-sensitive filters for the extended-emission structures used in the medium-band selection of \citet{zhu_systematic_2025}; F150W traces the stellar continuum in the Cosmic Ember. The detection image is smoothed with a Gaussian kernel with $\sigma=1$ pixel to suppress isolated noise peaks before thresholding. We estimate the local background with sigma-clipped statistics and define a source mask from pixels above a $2\sigma$ threshold. This threshold is used to define the spatial footprint for fitting rather than to assign formal detection significance to individual physical-property measurements; after masking, the selected pixels are typically detected in many bands, with median counts of $N_{\rm band}({\rm S/N}>2)=11$, 10, and 11 for the Cosmic Cigar, the Cosmic Rose, and the Cosmic Ember, respectively. This mask is dilated by one pixel to include low-surface-brightness outskirts.

We manually mask interlopers and adjacent sources within the cutout. In the Cosmic Rose, a low-redshift galaxy near the southern edge of the stamp is removed with a rectangular mask. The final fitted-pixel catalog includes pixels that pass this source-mask selection and have valid flux uncertainties in at least three filters. This yields 4606 fitted pixels for the Cosmic Cigar, 4532 for the Cosmic Rose, and 7340 for the Cosmic Ember. After requiring finite posterior medians for stellar mass and ${\rm SFR}_0$, the corresponding valid-pixel counts are 4578, 4475, and 7286.

\subsection{Spatially Resolved SED Fitting}\label{subsec:sed}

We fit the multi-band NIRCam photometry of each selected pixel independently. The goal is to map relative spatial variations in stellar-population and ISM-related quantities within each source. Each input pixel is assigned the systemic redshift measured from the spectra described in Section~\ref{sec:data}. Flux uncertainties are taken from the image error products. We caution that, because the images are PSF matched, neighboring pixels have correlated fluxes and are not statistically independent.

We follow a similar approach to that used in \citet{zhu_clumps_2026}. The SED modeling uses the {\tt Prospector} framework \citep{johnson_stellar_2021}, accelerated with the {\tt Parrot} artificial neural-network emulator \citep{mathews_as_2023}. The underlying stellar-population modeling is based on FSPS \citep{conroy_propagation_2009,conroy_propagation_2010}. We adopt a Chabrier initial mass function \citep{chabrier_galactic_2003} and a flexible non-parametric star-formation history similar to the Prospector-$\alpha$ framework \citep{leja_how_2019}. Our non-parametric SFH has seven age bins. The original bin edges are logarithmically spaced from $\log_{10}(t/{\rm yr})=7.1295$ to $\log_{10}(0.9\,t_{\rm univ}/{\rm yr})$, with a final bin extending to $\log_{10}(t_{\rm univ}/{\rm yr})$, where $t_{\rm univ}$ is the age of the Universe at the galaxy redshift and $t$ is measured as lookback time from that redshift. We modify the two youngest edges by replacing the first edge, $10^{7.1295}$~yr, with zero lookback time. The youngest bin, ${\rm SFR}_0$, therefore spans 0--$\sim$30~Myr, while ${\rm SFR}_1$ spans $\sim$30~Myr to the next logarithmically spaced edge, approximately 65~Myr for these targets. The model includes stellar mass, stellar metallicity, gas-phase metallicity, nebular emission, and dust attenuation. The same model form, priors, and fitting setup are applied to every pixel within a source, but the physical parameters are fit independently for each pixel. The maps are therefore used mainly for differential comparisons within each source. We sample the posterior with {\tt Nautilus}, a neural-network-assisted nested sampler \citep{lange_nautilus_2023}.

The fits return posterior constraints on the physical parameters for each valid pixel. In this paper we focus on five quantities: stellar-mass surface density, $\log(\Sigma_\star/M_\odot\,{\rm kpc}^{-2})$; specific star-formation rate, $\log({\rm sSFR}/{\rm yr}^{-1})$; the Prospector dust parameter \texttt{Dust2}; gas-phase metallicity, $\log(Z_{\rm gas}/Z_\odot)$; and the recent-to-past star-formation ratio $\log({\rm SFR}_0/{\rm SFR}_1)$. The \texttt{Dust2} parameter is the dimensionless optical-depth normalization of the diffuse dust component in the adopted \textsc{Prospector} attenuation model. The sSFR map uses the SFR in the youngest non-parametric SFH bin, corresponding to the most recent 30 Myr, divided by the fitted stellar mass. The gas-phase metallicity is constrained through the nebular-emission model and the sensitivity of the NIRCam filters, especially medium bands, to strong rest-frame optical line complexes. It should therefore be interpreted as model-dependent and mainly useful for relative spatial comparisons. In our setup, ${\rm SFR}_0$ and ${\rm SFR}_1$ correspond to 0--30 Myr and 30--65 Myr. Positive $\log({\rm SFR}_0/{\rm SFR}_1)$ indicates a rising recent SFH, while negative values indicate that the most recent 30 Myr bin is weaker than the preceding 30--65 Myr bin. As an external benchmark for the workflow, Appendix~\ref{app:ganifs_gs5001} applies the same \textsc{PhotoIFU} procedure to GS5001, a source with published NIRSpec IFU maps from GA-NIFS.

We use posterior medians to construct the maps shown in Figure~\ref{fig:sed_maps}. The spatial trends should be interpreted as differential trends within each target. PSF matching to F444W gives consistent colors across filters, but it also mixes neighboring structures. Compact gradients and radial patterns in the maps should therefore not be interpreted as deconvolved physical structures. This limitation mainly affects the spatial scale and absolute interpretation of individual pixels. The region comparisons below use the same PSF-matched data and the same SED-fitting assumptions for all pixels within a given source.

The physical-property maps in Figure~\ref{fig:sed_maps} are shown only for fitted pixels with finite posterior medians for the displayed quantity. A one-pass interpolation is applied only for display: isolated non-finite pixels whose eight neighboring pixels are finite are filled by the neighbor median. Non-finite pixels are not used in the subsequent analyses.

\section{Results}\label{sec:results}

\subsection{Resolved SED Maps}\label{subsec:maps}

Figure~\ref{fig:sed_maps} shows the basic result of treating NIRCam as a photometric IFU. The three systems show spatial variations in their SED-derived stellar-population and ISM-related properties. The maps of $\log(\Sigma_\star/M_\odot\,{\rm kpc}^{-2})$, $\log({\rm sSFR}/{\rm yr}^{-1})$, \texttt{Dust2}, $\log(Z_{\rm gas}/Z_\odot)$, and $\log({\rm SFR}_0/{\rm SFR}_1)$ go beyond the RGB light distribution by separating color, dust, metallicity, and recent-SFH differences.

For the Cosmic Cigar, the extended ionized-emission axis overlaps regions with distinct dust and recent-SFH properties. For the Cosmic Rose, the maps show multiple clumps or components with different SED properties, and the extended [O\,\textsc{iii}] emission may be associated with this clumpy morphology. For the Cosmic Ember, the maps show dust, sSFR, gas-phase metallicity, and recent-to-past SFR variations in a source selected through neutral absorption. Some nearly radial structure in the Cosmic Ember is shaped by the F444W-matched PSF and the strong central light concentration. We will explore the feasibility of de-convolving the PSF pattern in a future work.

\begin{figure*}[!ht]
\centering
\includegraphics[width=\textwidth]{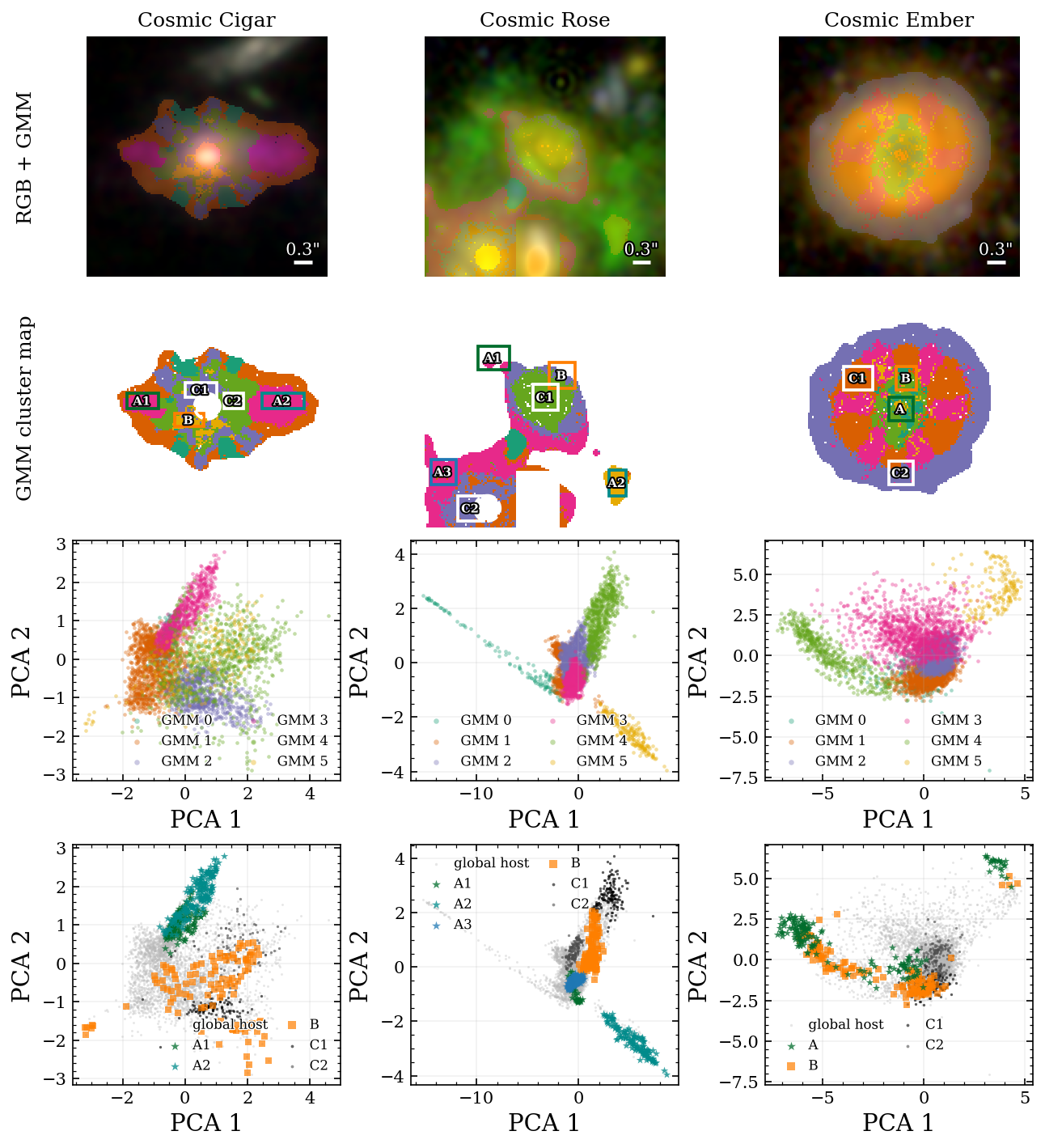}
\caption{Descriptive SED-property clustering of the resolved NIRCam pixels. Columns show the sources. From top to bottom, the rows show the RGB image with GMM clusters overlaid, the image-plane GMM cluster map using six components, the PCA projection colored by GMM cluster, and the same PCA projection with the manually selected gas-selected regions overlaid. The clustering uses only SED-derived physical properties and does not include pixel coordinates, RGB information, morphology, or manual region labels. Coherent structures in the image-plane cluster maps indicate spatially organized regions of pixels with similar fitted properties.}
\label{fig:gmm_pca}
\end{figure*}

\subsection{SED-property clustering}\label{subsec:cluster_results}

We use clustering as a descriptive check on whether pixels with similar fitted SED properties map back to coherent image-plane structures. This test does not identify feedback by itself. It asks whether the resolved SED maps contain spatially organized regions beyond single-filter morphology.

For each valid fitted pixel, we construct a feature vector using the posterior-median values of $\log(M_\star/M_\odot)$, $\log({\rm sSFR}/{\rm yr}^{-1})$, \texttt{Dust2}, $\log(Z_{\rm gas}/Z_\odot)$, and $\log({\rm SFR}_0/{\rm SFR}_1)$. Pixels with missing or non-finite values in any of these quantities are excluded from the clustering. The clustering input does not include pixel position, radius, RGB color, morphology, or any manually defined region label. The features are standardized before clustering so that no parameter dominates because of its numerical range.

We use principal component analysis (PCA) to visualize the multi-parameter pixel distribution in two dimensions. The PCA axes are linear combinations of the input SED quantities and are used here for visualization. We then use Gaussian mixture modeling (GMM) to identify groups of pixels with similar locations in the full SED-property space \citep{pedregosa_scikit-learn_2011}. In Figure~\ref{fig:gmm_pca}, we use six GMM components for each target. This fixed choice separates the main pixel groups while keeping the maps readable across the three galaxies. The cluster labels are reordered by increasing median $\log({\rm SFR}_0/{\rm SFR}_1)$, or by increasing $\log({\rm sSFR}/{\rm yr}^{-1})$ when the SFH-ratio map is unavailable.

Figure~\ref{fig:gmm_pca} shows that the GMM components project back onto coherent image-plane structures. The top two rows show this mapping directly in the image plane, while the lower two rows show the same pixels in PCA space with either GMM labels or manual gas-selected regions overlaid. The spatial coherence is not imposed by the clustering because the clustering contains no spatial information. It appears because neighboring pixels have similar fitted SED properties. Our GMM clustering results may hint the localized impact of feedback.

\begin{figure*}[!ht]
\centering
\includegraphics[width=\textwidth]{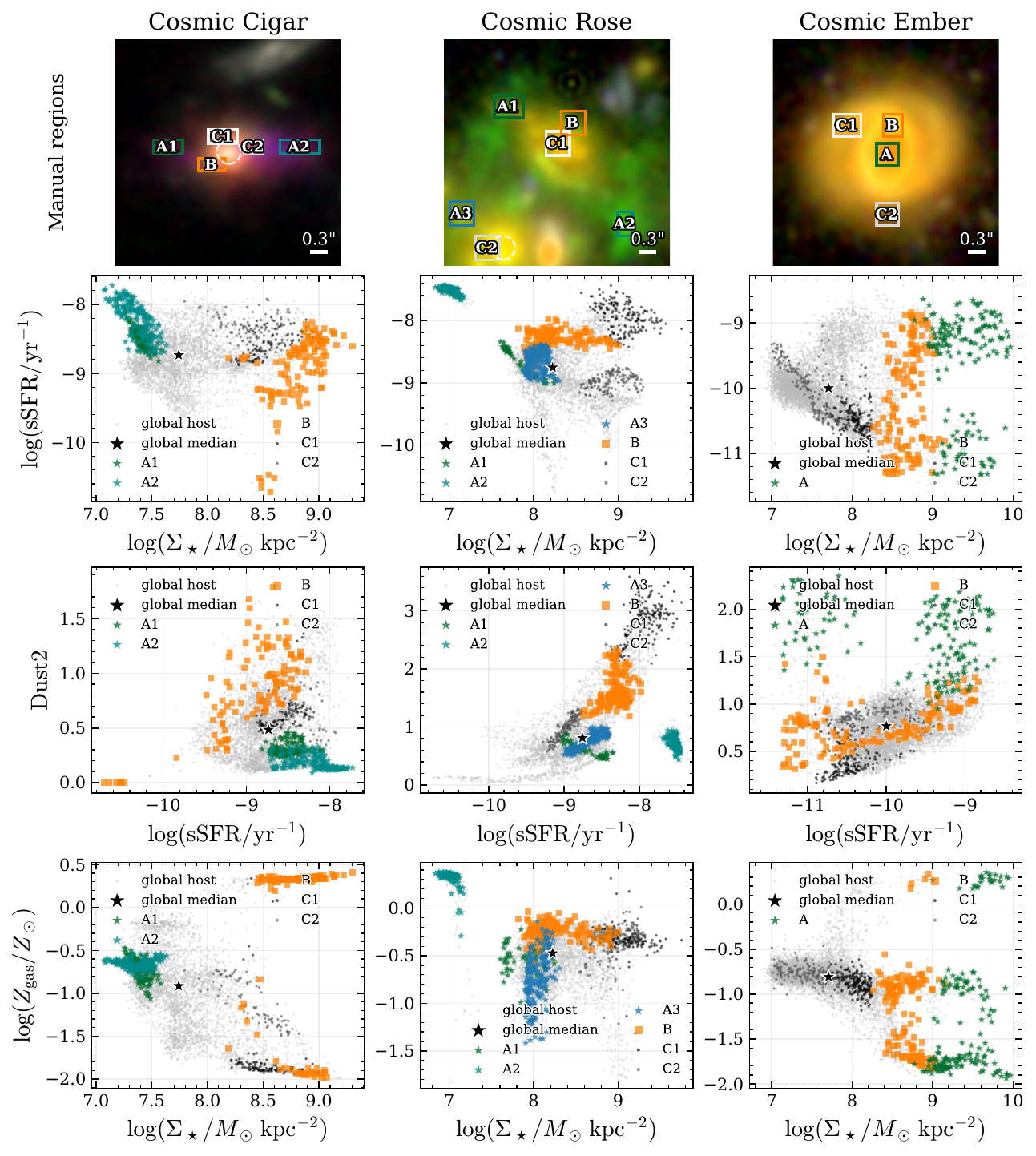}
\caption{Manual gas-selected regions and their locations in resolved SED-property planes. Columns show the sources, and rows show the image-plane manual regions and the SED-property projections: $\log({\rm sSFR}/{\rm yr}^{-1})$ versus $\log(\Sigma_\star/M_\odot\,{\rm kpc}^{-2})$, \texttt{Dust2} versus $\log({\rm sSFR}/{\rm yr}^{-1})$, and $\log(Z_{\rm gas}/Z_\odot)$ versus $\log(\Sigma_\star/M_\odot\,{\rm kpc}^{-2})$. Global host pixels are shown as the light background population, while manually defined regions are shown with colored symbols. The black star marks the global host median of the fitted pixels in each panel.}
\label{fig:manual_correlations}
\end{figure*}

\subsection{SED Properties in Manually Selected Regions}\label{subsec:phase_space}

We next compare the SED-only clustering with a fixed image-plane region test. The manually selected regions in Figure~\ref{fig:manual_correlations} are apertures chosen from the RGB images, the line-sensitive medium-band morphology, and the known spectroscopic gas signatures. The regions are defined before comparing their SED-derived properties. They are not selected from the SED-property maps, the GMM clusters, or the PCA projection. The clustering asks whether pixels with similar fitted SED properties are spatially organized. The fixed-region test asks whether pixels in predefined regions, especially regions close to the gas signatures, differ from the full host distribution. The rectangular boundaries shown in Figure~\ref{fig:manual_correlations} are the apertures used in the analysis; their sizes were chosen to isolate the identified structures while minimizing overlap between adjacent regions.

For the Cosmic Cigar, the A regions sample the extended ionized-emission bicone, while the B and C regions sample off-axis or local host comparison regions. The central AGN/core region is excluded from the region-based comparisons. For the Cosmic Rose, the A, B, and C regions sample the clumpy system and its extended [O\,\textsc{iii}] structure. The likely AGN/core region associated with the lower-left, southeastern component is excluded. For the Cosmic Ember, the Na\,\textsc{D} absorption is measured in the SMILES MSA spectrum, so the A region tests the resolved SED properties along the absorption sight line rather than a transverse map of the neutral outflow. The approximate MSA shutter location is shown in Figure~\ref{fig:rgb_spectra}. The B and C regions sample surrounding host structures.

The A-region centroids span projected separations of $9.0$--$10.6~{\rm kpc}$ in the Cosmic Cigar, $8.6$--$14.7~{\rm kpc}$ in the Cosmic Rose, and $0.4~{\rm kpc}$ in the Cosmic Ember from the adopted source centers.

Figure~\ref{fig:manual_correlations} shows that these visible structures occupy restricted regions within the full host SED-property distribution. The Cosmic Cigar is the clearest ionized-outflow case: its A regions trace the extended axis and sit away from much of the host population, especially in dust and recent-SFH space. The Cosmic Rose is more complex. The A1/A2/B/C1 group and the A3/C2 group are separated both spatially and in SED-property space, consistent with a clumpy multi-component system where the lower-left, southeastern component is physically distinct. The Cosmic Ember shows that the same approach is useful for a neutral-outflow fossil system. Its selected regions separate in SED-property space even without a bright extended ionized-emission map.

\begin{figure*}[!ht]
\centering
\includegraphics[width=\textwidth]{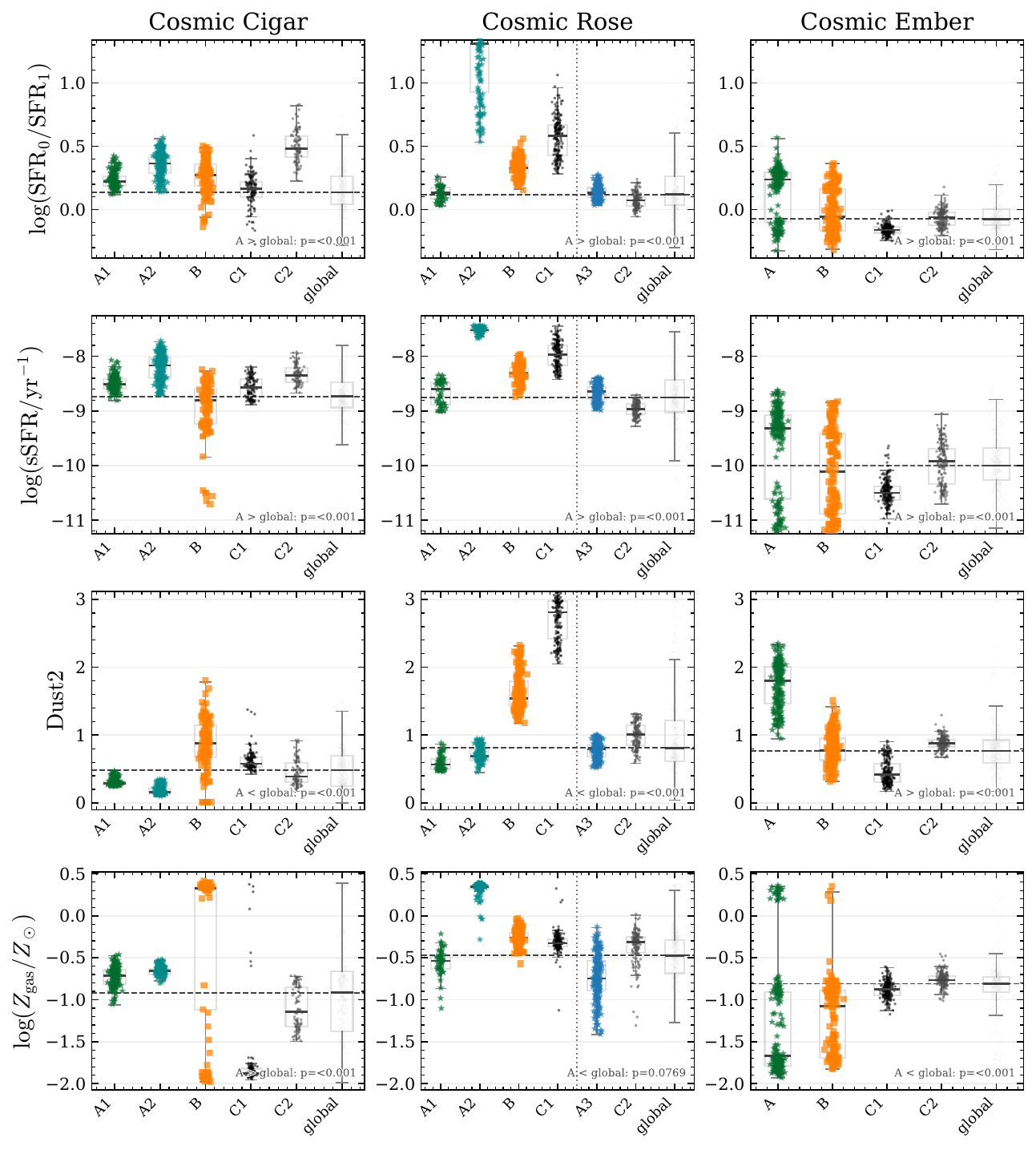}
\caption{Region-by-region distributions for the manual regions. Columns show the sources, and rows show $\log({\rm SFR}_0/{\rm SFR}_1)$, $\log({\rm sSFR}/{\rm yr}^{-1})$, \texttt{Dust2}, and $\log(Z_{\rm gas}/Z_\odot)$. The global reference includes all valid host SED pixels after applying the same non-finite-pixel and core-exclusion masks used for the region analysis, not only pixels inside manual masks. The dashed line marks the global median. The p-value in each panel compares the pooled A-region pixels with the remaining valid host pixels outside the pooled A regions using a two-sided Mann--Whitney U test; the direction label indicates whether the A-region median is higher or lower than the global median. Because the pixels are PSF-correlated, the p-values are used only as descriptive distribution-separation diagnostics and should not be interpreted as independent-pixel significances. For the Cosmic Rose, the vertical divider separates the likely A1/A2/B/C1 group from the lower or secondary A3/C2 component.}
\label{fig:manual_comparisons}
\end{figure*}

\subsection{Region-by-region Distribution Comparisons}\label{subsec:region_distributions}

Figure~\ref{fig:manual_comparisons} quantifies the region differences. For each target and property, we compare the region distributions against the global host distribution. Here ``global'' means all valid fitted host SED pixels after applying the same non-finite-pixel and target-level core-exclusion masks used for the region analysis. It is not restricted to pixels inside the manual boxes. The dashed horizontal line marks the median of this global host distribution.

We use a two-sided Mann--Whitney U test comparing the pooled A-region pixels with the remaining valid host pixels outside the pooled A regions. This non-parametric test asks whether the two samples tend to occupy different value ranges, without assuming a Gaussian distribution. The p-value gives the probability of obtaining a separation at least this strong if the two samples were drawn from the same parent distribution. The direction label, for example A $>$ global or A $<$ global, is assigned from the median offset. Because the pixels are PSF-correlated, these p-values are used only as descriptive distribution-separation diagnostics and should not be interpreted as independent-pixel significances.

The main empirical result is that the A regions are often offset from the global host distribution. In the systems with extended ionized emission, the offsets commonly involve lower \texttt{Dust2} values. This is consistent with extended ionized emission being observed along lower-attenuation, possibly lower-column-density pathways through the host. Such pathways can allow ionizing photons to travel farther from their source and can make extended line emission easier to observe. In the Cosmic Ember, the strongest offsets involve recent star formation and dust attenuation. These offsets may mark a broader recent disturbance, such as gas inflow, compaction, merger-related activity, or a faded AGN phase, that is also connected to the observed Na\,\textsc{D} outflow.

Quantitatively, the median A-region \texttt{Dust2} offsets relative to the global host are $\Delta\texttt{Dust2}=-0.24$ for the Cosmic Cigar and $\Delta\texttt{Dust2}=-0.07$ for the Cosmic Rose. In the Cosmic Ember, the largest A-region offsets are $\Delta\texttt{Dust2}=+1.04$, $\Delta\log(Z_{\rm gas}/Z_\odot)=-0.86$, and $\Delta\log({\rm sSFR}/{\rm yr}^{-1})=+0.68$.

\section{Discussion}\label{sec:discussion}

\subsection{A photometric-IFU view of feedback}
\label{subsec:phot_ifu_feedback}

A main result of this work is that gas-selected regions occupy distinct parts of the resolved SED-property distribution in all three systems studied here. Despite spanning an AGN-driven ionized outflow, a clumpy galaxy with extended ionized emission, and a post-starburst galaxy with a powerful neutral outflow, each system shows measurable differences between the gas-associated regions and the broader host galaxy. These differences involve dust attenuation, recent star formation, metallicity, and stellar-population structure. While they do not uniquely identify the physical origin of the gas, they demonstrate that feedback signatures are spatially associated with distinct host-galaxy environments.

The key advance of \textsc{PhotoIFU} is that it links gas signatures to resolved stellar-population properties using imaging alone. Traditional feedback studies begin with spectroscopy, which measures gas kinematics, densities, and ionization diagnostics. These observations reveal where the gas is and how it moves, but they do not directly connect the gas to the resolved stellar populations across the host galaxy. By treating deep multi-band NIRCam imaging as a low-resolution photometric IFU, \textsc{PhotoIFU} provides galaxy-wide maps of stellar mass, recent star formation, dust attenuation, and nebular properties that can be compared directly with independently identified gas structures.

The unsupervised clustering provides an independent consistency check on this interpretation. Because the GMM analysis uses only SED-derived physical properties and excludes spatial coordinates, the coherent image-plane structures are not imposed by the algorithm. Instead, neighboring pixels naturally cluster because they share similar stellar-population and ISM-related properties. The manually defined gas-selected apertures then provide a complementary test tied directly to the observed spectroscopic or line-emission features.

This imaging-based approach is not intended to replace spatially resolved spectroscopy. NIRSpec IFU observations remain essential for measuring gas velocities, densities, ionization states, and chemical abundances. Instead, \textsc{PhotoIFU} complements spectroscopy by providing full spatial coverage and enabling resolved stellar-population analyses for systems where complete IFU mapping is impractical. It also offers a practical way to prioritize future spectroscopic follow-up by identifying regions with the strongest physical contrasts within galaxies.

\subsection{Dust-poor channels and extended ionized emission}
\label{subsec:dust_channels}

For the systems with extended ionized emission, the most suggestive trend is that the gas-selected regions are often less dusty than the full host distribution. This does not require feedback to remove all dust along the outflow axis, and dust alone cannot determine the observed emission. The result is consistent with extended ionized emission being preferentially observed along dust-poor, low-column-density pathways through the host galaxy. Such pathways are expected in a porous ISM, where compact starbursts and radiation-driven gas dispersal can produce anisotropic LyC escape \citep[e.g.,][]{menon_bursts_2025}.

In this geometry, outflows, radiation pressure, stellar feedback, or AGN activity can clear or illuminate a small number of favorable sight lines, allowing ionizing photons to propagate farther from their sources and power extended line emission. Recent JWST observations likewise emphasize the importance of dust geometry for ionizing-photon escape \citep{ji_importance_2025}. Our maps provide a resolved view of candidate low-attenuation pathways in systems with extended ionized emission \citep[see also][]{peng_direct_2025}.

The Cosmic Cigar provides the clearest example. It contains a broad-line AGN and an extended ionized-emission structure. The regions along the extended axis are distinct from the full host in dust and recent-SFH space, consistent with an outflow or radiation channel passing through a lower-attenuation part of the galaxy. In this case, the NIRCam maps help connect the known gas signature to the host structure that may regulate how radiation and ionized gas escape.

The Cosmic Rose shows a more complicated version of the same problem. Its very extended [O\,\textsc{iii}] emission sits in a clumpy, multi-component system, and the lower-left, southeastern component is likely associated with an AGN as indicated by the X-ray and MIRI detection (J.\ Lyu et al. in prep). The SED maps and SED-property projections show that the different components are physically distinct. Such component-to-component differences are expected in clumpy high-redshift systems, where individual clumps and diffuse regions can show different attenuation properties \citep{zhu_clumps_2026,nakazato_clump-scale_2026,markov_resolving_2026}. Moreover, our results show that the extended line emission cannot be interpreted as a single smooth halo around one simple host. It is more likely tracing a combination of feedback, radiation transport, and the multi-component structure of the system. This is precisely where the photometric-IFU approach is useful: before assigning a physical origin to the extended gas, it separates the stellar-population and dust structure of the components that may be contributing to the observed emission.

\begin{figure*}[!ht]
\centering
\includegraphics[width=0.9\textwidth]{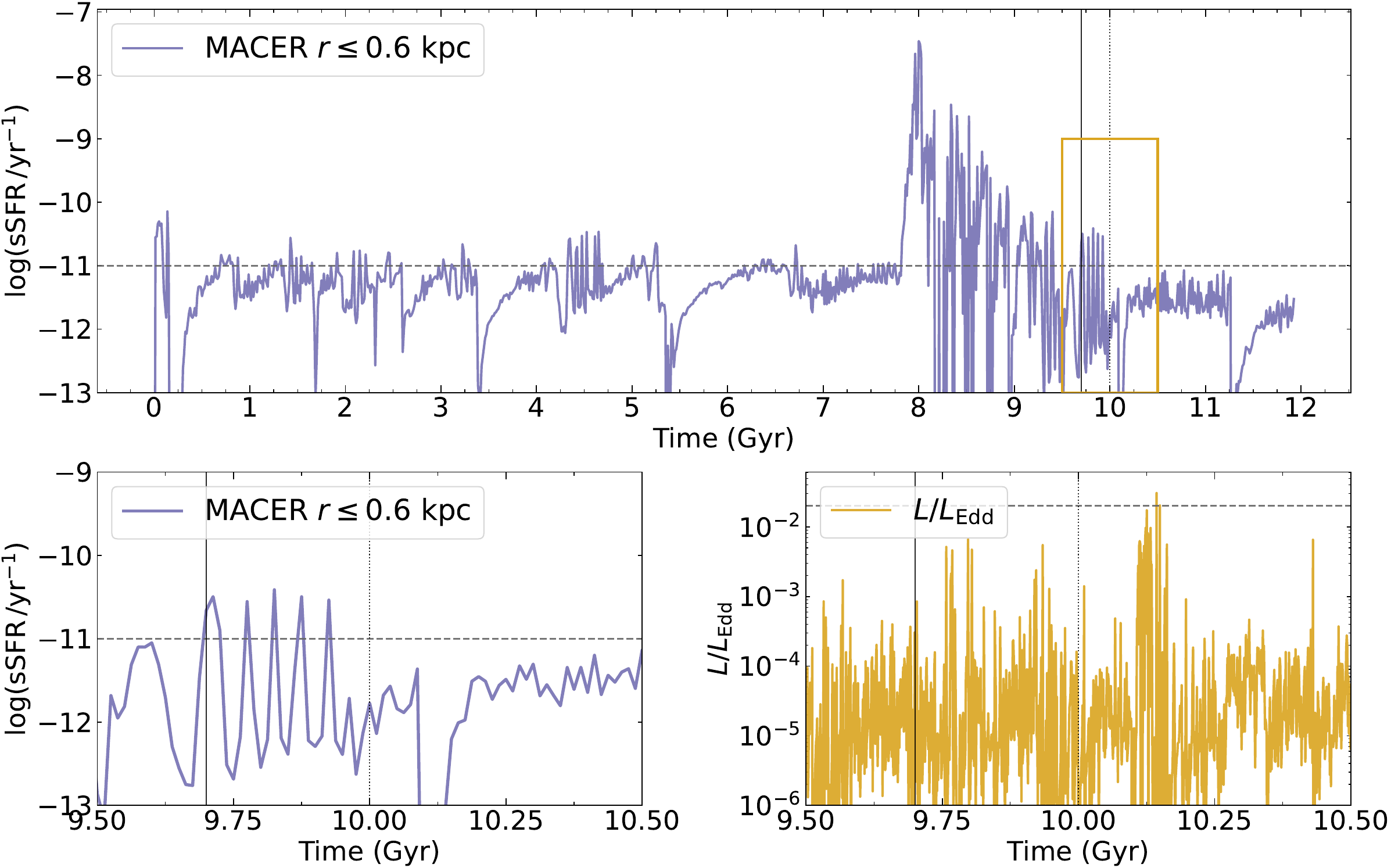}
\caption{MACER fiducial run. \emph{Top:} central ($r\leq0.6$~kpc) $\log({\rm sSFR})$ versus time. The golden box marks the interval shown in the lower panels. \emph{Bottom left:} zoom to $t=9.5$--10.5~Gyr. \emph{Bottom right:} black-hole Eddington ratio over the same interval. Solid and dotted vertical lines mark $t=9.70$ and 10.0~Gyr, corresponding to the rejuvenation and fossil-outflow comparison epochs discussed in the text.}
\label{fig:macer_ssfr}
\end{figure*}




\subsection{A fossil neutral outflow in the Cosmic Ember}
\label{subsec:cosmic_ember}

Because the Cosmic Ember has a powerful Na\,\textsc{D} outflow but a low present-day SFR and no clear luminous AGN \citep{sun_extreme_2025,zhu_smiles_2026}, it raises two linked questions: whether the outflow was launched in an earlier energetic phase, and whether the compact central component seen in the \textsc{PhotoIFU} maps and morphology fits (Appendix~\ref{app:ember_morphology}) traces later gas return, compact star formation, or residual activity in the post-starburst host.

To explore whether a faded AGN-driven launch and subsequent central gas return can coexist in a post-starburst host, we compare the Cosmic Ember with an illustrative axisymmetric MACER simulation of a massive disk galaxy with self-consistent black-hole accretion and two-mode AGN feedback \citep{yuan_active_2018,zou_systematic_2026}. MACER evolves a single gaseous disk with cosmological inflows, radiative quasar winds, and hot jets coupled to the ISM. Although the model is two-dimensional and cannot reproduce pixel-level NIRCam SED maps, it captures the coupled evolution of AGN-driven mass ejection, global star-formation decline, and localized nuclear gas return that is relevant for post-starburst rejuvenation.

In the fiducial MACER run (\(\nu=0.55\)\footnote{\(\nu\) is the spatially uniform kinematic viscosity in the MACER simulations \citep{zou_systematic_2026}, used to parameterize the efficiency of angular-momentum transport; larger \(\nu\) implies more efficient transport.}), a cold-filament-driven starburst at $t\approx8.0$~Gyr reaches $\mathrm{SFR}\sim2.2\times10^{2}\,M_\odot\,\mathrm{yr}^{-1}$, followed by a strong AGN outflow episode at $t\approx8.4$~Gyr with $\dot M_{\rm out}\sim6.5\times10^{1}\,M_\odot\,\mathrm{yr}^{-1}$ in the wind/jet tracer field ($\log\dot M_{\rm out}\approx1.8$). The predicted AGN duty cycle of $\sim0.49\%$ is consistent with observations \citep{zou_systematic_2026}. The galaxy is then quenched on a $\sim1$~Gyr timescale by the enhanced AGN activity. By $t\sim9.5$--10.5~Gyr the model enters a low-$L/L_{\rm Edd}$ state in which the neutral-wind kinematic signature can outlast the luminous AGN phase, while the central gas reservoir can begin to rebuild. Figure~\ref{fig:macer_ssfr} summarizes the late-time sSFR and accretion evolution. 
The corresponding slice maps and mass-flux diagnostic are shown in Appendix~\ref{app:macer_comparison}.
We note that the simulation time is used only to identify an evolutionary phase within the model and is not necessarily matched to the cosmic age of the Cosmic Ember.

Figure~\ref{fig:macer_ssfr} shows the long-term decline of the central ($r\leq0.6$~kpc; comparable to the core size in our SED map) $\log({\rm sSFR})$ after the burst and a zoom to $t=9.5$--10.5~Gyr, together with the Eddington ratio. Vertical lines mark $t=9.70$~Gyr, when the central sSFR rebounds above the preceding post-starburst baseline, and $t=10.0$~Gyr, when fossil wind/jet material is still present in the inner kiloparsec. This sequence provides a qualitative match to the Cosmic Ember, where the powerful Na\,\textsc{D} outflow reported above likely records an earlier energetic phase while the present-day host shows localized recent star formation with no clear ongoing luminous AGN activity \citep{sun_extreme_2025}.

At $t=9.70$~Gyr the simulation develops a compact central rejuvenation episode (Figure~\ref{fig:macer_reju_slices}; Appendix~\ref{app:macer_comparison}). The radial-velocity panel reveals outward-moving gas in the inner few kiloparsecs, while the density and temperature panels show that relatively cold gas remains in the central disk. The abundance panels indicate that gas with initial stellar abundance tracers is still present and has not been fully expelled by the earlier quenching outflow. We interpret this as residual ISM gas that remained stable after the AGN-driven ejection episode and can fall back toward the center once $L/L_{\rm Edd}$ drops, producing the localized sSFR rise seen in Figure~\ref{fig:macer_ssfr}. Within $r\leq2$~kpc, wind/jet-traced outflowing gas still has $v_{r,\,90}\approx7.3\times10^{2}$~km~s$^{-1}$, comparable in order of magnitude to the Na\,\textsc{D}-inferred outflow speed in the Cosmic Ember ($v_{\rm out}\sim828$~km~s$^{-1}$), while the instantaneous shell mass flux at $r=1$~kpc remains low ($\dot M_{\rm tracer}\approx10^{-3}\,M_\odot\,\mathrm{yr}^{-1}$).

At $t=10.0$~Gyr the simulation enters a clearer fossil-outflow stage (Figures~\ref{fig:macer_outflow_r1kpc} and \ref{fig:macer_fossil_slices}; Appendix~\ref{app:macer_comparison}). Legacy AGN-driven material is still visible in $v_r$, density, and the wind/jet abundance panel within the inner few kiloparsecs, even though $L/L_{\rm Edd}$ is low. Within $r\leq2$~kpc, the 90th percentile radial velocity of wind/jet-traced outflowing gas reaches $v_{r,\,90}\approx9.9\times10^{2}$~km~s$^{-1}$, again comparable to the Na\,\textsc{D} measurement in the Cosmic Ember. We therefore emphasize a primarily kinematic comparison at this fossil stage, rather than a direct match to the Na\,\textsc{D}-inferred effective mass-loss rate \citep{sun_extreme_2025}. Together, the simulation suggests that fossil neutral-wind kinematics and a compact central rejuvenation episode can coexist in a post-starburst host, providing a physically plausible scenario for the spatial structure seen in the \textsc{PhotoIFU} maps.

\subsection{Connecting gas signatures to resolved host structure}
\label{subsec:resolved_imprints}

The broader question motivating this paper is whether galactic outflows are associated with resolved host-galaxy differences. For the three systems studied here, regions associated with known gas signatures occupy distinct parts of the host SED distribution, and this behavior appears in both the SED-only clustering and the manually selected gas-selected apertures. A physical framework connecting such resolved structure to compaction and quenching has been proposed in cosmological zoom-in simulations of $z\sim1$--3 galaxies with stellar masses comparable to our targets ($M_\star\sim10^{10}$--$10^{11}\,M_\odot$). Intense gas inflow through mergers, counter-rotating streams, or recycled gas drives dissipative ``wet'' compaction, followed by a burst of central star formation and subsequent inside-out quenching through gas depletion by star formation and stellar, supernova, or AGN feedback \citep{zolotov_compaction_2015,tacchella_evolution_2016,tacchella_confinement_2016,ceverino_effects_2023}.

\citet{zolotov_compaction_2015} show that long-term quenching requires the host halo to exceed a threshold mass sufficient to sustain a stable virial shock and a hot circumgalactic medium that suppresses further cold-gas supply \citep{dekel_galaxy_2006,dekel_cold_2009}. Below this threshold, fresh gas can continue to reach the center, allowing compaction and quenching episodes to recur. Based on their stellar masses and redshifts, our targets may inhabit halos of order $10^{12}\,M_\odot$ according to the mean stellar-to-halo mass relation of \citet{behroozi_universemachine_2019}, although the halo masses of individual systems remain uncertain. This places them near the regime where stable virial shocks may develop, while cold streams can still penetrate massive hot halos at high redshift \citep{dekel_galaxy_2006,dekel_cold_2009}. The resolved maps cannot distinguish fresh cosmological inflow from recycled or fallback gas, but the latter provides one possible origin for the central star-formation enhancement, as illustrated for the Cosmic Ember in Section~\ref{subsec:cosmic_ember}.

Consistent with this picture, at higher redshift and lower stellar mass ($\log M_\star\lesssim9$ at $z\gtrsim4$), recent JWST studies report rapidly varying star-formation histories, often characterized using short-to-long-timescale SFR indicators such as ${\rm SFR}_{10\rm\,Myr}/{\rm SFR}_{100\rm\,Myr}$. These include candidate mini-quenching systems \citep{covelo-paz_systematic_2026,strait_extremely_2023,looser_recently_2024,looser_jades_2025,baker_zapped_2025,harikane_uv-luminous_2026} and rejuvenation events \citep{witten_rising_2025}, with similar behavior in cosmological zoom-in simulations \citep{dome_mini-quenching_2024,mcclymont_thesan-zoom_2025}.

These results support a picture in which feedback and host structure evolve together. Outflows propagate through an inhomogeneous ISM, with paths shaped by local dust column, gas density, stellar population, and recent star-formation history. The resulting SED differences are not universal: in some systems they appear as lower dust along an ionized-emission axis, while in others they appear as enhanced recent star formation near neutral gas. The common feature is that gas-selected regions are offset from the full host in ways that can be mapped with deep multi-band imaging.

Our interpretation remains subject to the limitations of resolved SED fitting. Pixel-level quantities depend on the assumed stellar-population model, dust prescription, nebular-emission treatment, priors, and possible AGN contribution. PSF matching to F444W mixes neighboring structures, so adjacent pixels are not independent, while projection effects, compact-source outshining, and uncertainty in the ionizing source of extended emission remain. The spatial extent of compact gradients should therefore be treated as an upper limit. The central result is differential: under the same imaging, PSF, and SED-fitting assumptions, the gas-selected regions occupy distinct parts of the host distribution. NIRSpec IFU observations remain the cleanest way to measure gas velocities, densities, ionization states, and abundances, but NIRCam \textsc{PhotoIFU} mapping can identify where the strongest physical contrasts occur, guide follow-up spectroscopy, and extend resolved analyses to larger samples where full IFU mapping is not feasible.

\section{Summary}\label{sec:summary}

We use deep JWST/NIRCam imaging as a low-resolution photometric IFU to ask whether gas signatures of feedback or outflows are associated with resolved host-galaxy differences. In all three systems studied here, regions selected from extended ionized emission or neutral-outflow geometry occupy distinct parts of the resolved SED-property distribution. This shows that gas-selected structures can be spatially associated with measurable differences in the stellar populations and ISM-related properties of their hosts.

The main results are:

\begin{enumerate}
    \item In the Cosmic Cigar, the regions along the extended ionized-emission axis are offset from the full host distribution, especially in dust attenuation and recent-SFH space. This supports a picture in which the ionized outflow or radiation field is associated with dust-poor, low-column-density pathways through the host.

    \item In the Cosmic Rose, the extended [O\,\textsc{iii}] emission is embedded in a clumpy, multi-component system whose components have different resolved SED properties. The gas-selected regions split into distinct host components, indicating that the large-scale ionized emission must be interpreted together with the internal structure of the system.

    \item In the Cosmic Ember (JADES/SMILES 206183), a post-starburst galaxy with a powerful Na\,\textsc{D} neutral outflow and no clear evidence for ongoing luminous AGN activity, the selected regions show enhanced recent star formation and dust structure. This may mark the aftermath of an earlier energetic phase or a broader recent disturbance connected to the observed outflow.

    \item An illustrative MACER simulation shows that a fossil outflow can persist after the luminous AGN phase while gas returns to the central region and produces a compact rejuvenation episode. The simulation provides a plausible physical scenario for our observations.

    \item More broadly, the resolved differences are consistent with simulation-based pictures in which gas inflow, compaction, feedback, quenching, and rejuvenation are coupled during galaxy evolution. Across the three systems, the SED-only GMM/PCA analysis identifies pixel groups that map back onto coherent image-plane structures, while the manually selected gas-associated regions remain offset from the full host distribution.
\end{enumerate}

The same approach will be useful in the Roman and Rubin eras, where deep wide-field imaging can identify candidate systems for spectroscopic follow-up and provide resolved or semi-resolved stellar-population constraints for samples that are too large for full IFU mapping.

\begin{acknowledgments}
We acknowledge support from the NIRCam Science Team contract to the University of Arizona, NAS5-02105. Y.Z. is also supported by JWST Program \#6434. Support for program \#6434 was provided by NASA through a grant from the Space Telescope Science Institute, which is operated by the Association of Universities for Research in Astronomy, Inc., under NASA contract NAS 5-03127.

C.C. acknowledges support from the JWST/NIRCam Science Team contract to the University of Arizona, NAS5-02105, and JWST Programs 3215 and 5015.
C.C.W. acknowledges support from NOIRLab, which is managed by the Association of Universities for Research in Astronomy (AURA) under a cooperative agreement with the National Science Foundation.
S.A. acknowledges support from the JWST Mid-Infrared Instrument (MIRI) Science Team Lead, grant 80NSSC18K0555, from NASA Goddard Space Flight Center to the University of Arizona.
Y.N. acknowledges a Flatiron Research Fellowship. The Flatiron Institute is a division of the Simons Foundation.
A.J.B.\ and J.C. acknowledge funding from the ``FirstGalaxies'' Advanced Grant from the European Research Council (ERC) under the European Union's Horizon 2020 research and innovation program (Grant agreement No. 789056).
P.G.P.-G. acknowledges support from grant PID2022-139567NB-I00 funded by Spanish Ministerio de Ciencia e Innovaci\'on MCIN/AEI/10.13039/501100011033, FEDER, UE.
B.R.P. acknowledges support from grant PID2024-158856NA-I00 funded by Spanish Ministerio de Ciencia e Innovaci\'on MCIN/AEI/10.13039/501100011033 and by ``ERDF A way of making Europe''.

This work is based on observations made with the NASA/ESA/CSA James Webb Space Telescope. The data were obtained from the Mikulski Archive for Space Telescopes at the Space Telescope Science Institute, which is operated by the Association of Universities for Research in Astronomy, Inc., under NASA contract NAS 5-03127 for JWST. All the JWST data used in this paper can be found in MAST: \dataset[https://doi.org/10.17909/8tdj-8n28]{https://doi.org/10.17909/8tdj-8n28}. The authors acknowledge the FRESCO (PI: P.~Oesch) team for developing their observing program with a zero-exclusive-access period.

This material is based upon High Performance Computing (HPC) resources supported by the University of Arizona TRIF, UITS, and Research, Innovation, and Impact (RII) and maintained by the UArizona Research Technologies department. This project made use of lux supercomputer at UC Santa Cruz, funded by NSF MRI grant AST 1828315. 
This manuscript benefited from grammar checking and proofreading using ChatGPT \citep{openai_chatgpt_2024}. The public version of \textsc{PhotoIFU} was developed with assistance from Codex, based on an internally developed codebase.

We respectfully acknowledge the University of Arizona is on the land and territories of Indigenous peoples. Today, Arizona is home to 22 federally recognized tribes, with Tucson being home to the O'odham and the Yaqui. The university strives to build sustainable relationships with sovereign Native Nations and Indigenous communities through education offerings, partnerships, and community service.

\end{acknowledgments}


\vspace{5mm}
\facilities{JWST, MAST}

\software{
    {\tt scikit-learn} \citep{pedregosa_scikit-learn_2011}
    }

\appendix

\begin{figure*}[!ht]
    \centering
    \includegraphics[width=\textwidth]{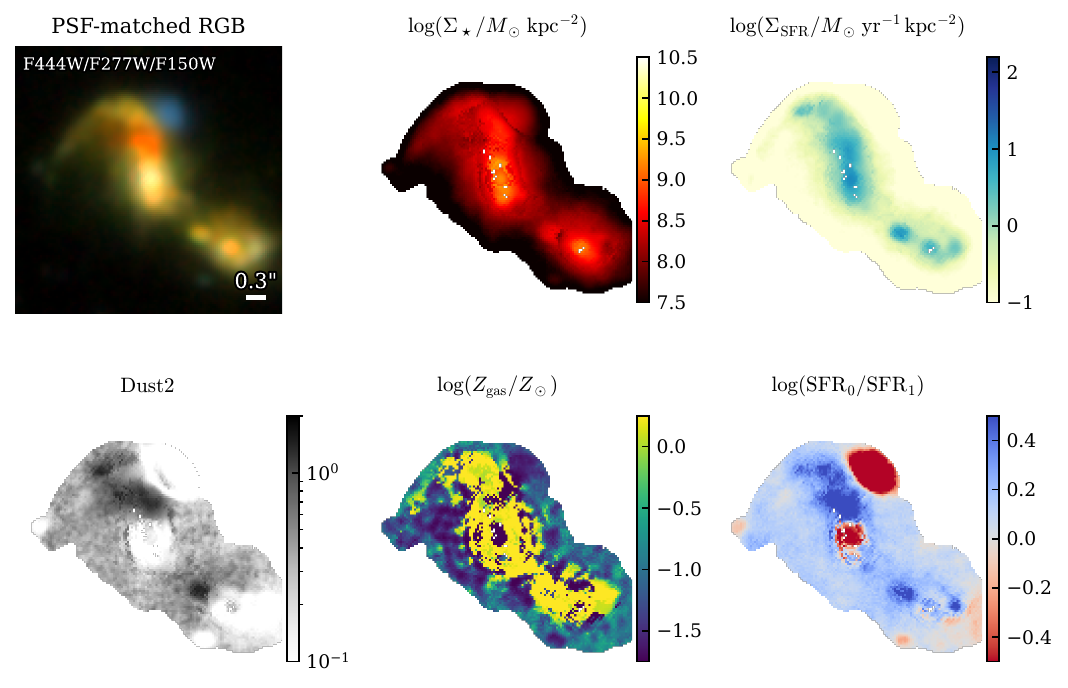}
    \caption{NIRCam \textsc{PhotoIFU} benchmark maps for the GA-NIFS source GS5001, corresponding to NIRCam ID 175485. The panels show the PSF-matched RGB image, projected stellar-mass surface density, projected SFR surface density, \texttt{Dust2}, gas-phase metallicity, and recent-to-past star-formation ratio. The mass and SFR maps are converted to projected physical surface densities using the $0.03''$ pixel scale and the angular-diameter distance of the source. These maps show broad spatial consistency with the NIRSpec IFU maps of \citet{lamperti_ga-nifs_2024}, especially in the dust and star-formation structure.}
    \label{fig:ganifs_gs5001_nircam}
\end{figure*}

\section{A NIRSpec IFU benchmark for \textsc{PhotoIFU}}
\label{app:ganifs_gs5001}

As an external benchmark for the \textsc{PhotoIFU} workflow, we apply the same procedure to GS5001, which corresponds to NIRCam ID 175485 and was observed with JWST/NIRSpec IFS by \citet{lamperti_ga-nifs_2024}. The benchmark is shown in Figure~\ref{fig:ganifs_gs5001_nircam}. The NIRCam-based maps recover stellar-population and ISM-related structure that is broadly consistent with the GA-NIFS view: the high-SFR regions trace the main galaxy and southern companion, the dust map highlights the obscured lane seen in the IFU attenuation map, and the relative metallicity structure is qualitatively similar to the spatial variations traced by the IFU line-ratio and metallicity maps. The gas-phase metallicity from \textsc{PhotoIFU} should be interpreted cautiously because it is inferred from low-resolution photometric SEDs and is sensitive to the nebular-emission model and filter sampling. The benchmark is therefore most useful as a spatial consistency check. It shows that \textsc{PhotoIFU} can recover the main resolved stellar-population and relative ISM-property patterns while extending the mapping over the full NIRCam footprint.

\begin{figure*}[!ht]
\centering
\includegraphics[width=\textwidth]{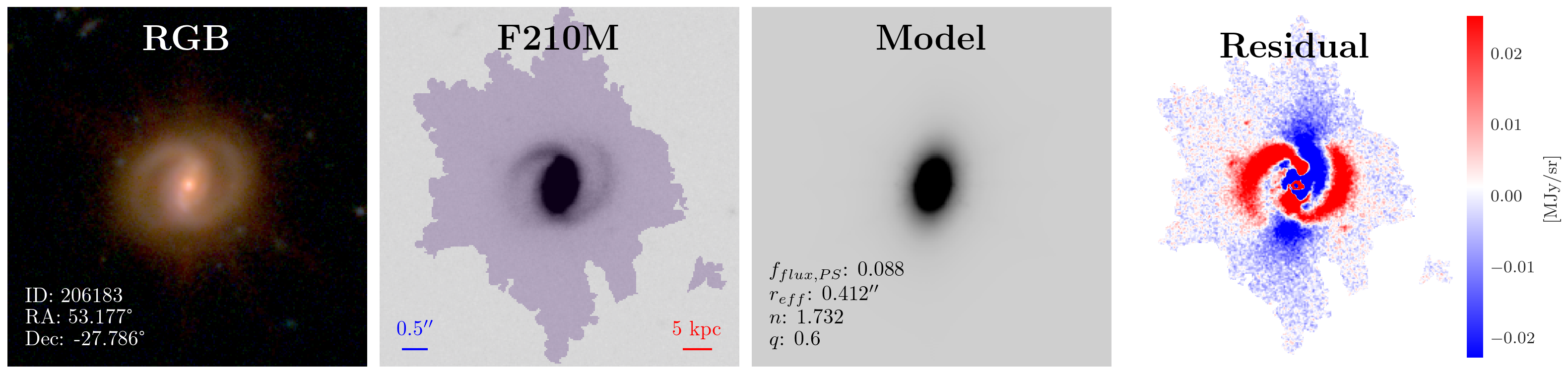}
\vspace{2mm}
\includegraphics[width=\textwidth]{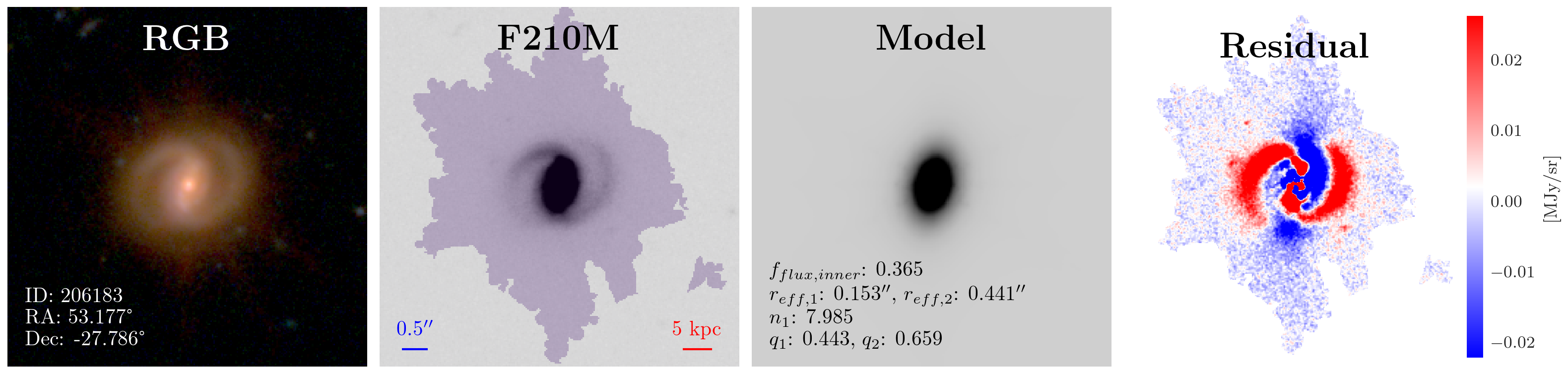}
\caption{Morphological fits to the Cosmic Ember F210M image. The RGB (F444W/F200W/F090W) and F210M panels are duplicated on both the top and bottom rows of the figure. Top: S\'ersic+PSF model. The effective radius, S\'ersic index $n$, and axis ratio $q$ of the galaxy are reported in the Model panel. In that panel, $f_{flux, PS}$ represents the fraction of the total image flux that can be attributed to the modeled central point source. Bottom: Two-S\'ersic model. The structural parameters for each S\'ersic component are reported in the Model panel; parameters with subscript 1 correspond to the inner component, while those with subscript 2 correspond to the outer component. The S\'ersic index of the outer component is fixed to $n_2=1$ in the modeling and is therefore not reported in the figure. Finally, $f_{flux, inner}$ indicates the fraction of the total flux of the image attributed to the inner component. Both fits capture the strong central concentration but leave structured residuals, indicating that the central component is not fully described by a simple unresolved PSF or by a smooth bulge-like model. The morphology indicates a bright compact central component with surrounding structured emission. Together with the lack of clear ongoing luminous AGN signatures from the spectra and MIRI constraints, this supports the interpretation of a compact rejuvenated or star-forming central region embedded in an extended post-starburst host.}
\label{fig:ember_morphology}
\end{figure*}

\section{Morphological fits to the Cosmic Ember}\label{app:ember_morphology}

Figure~\ref{fig:ember_morphology} shows simple morphological fits to the Cosmic Ember F210M image. The purpose of this test is not to build a full structural model of the galaxy, but to check whether the central light concentration can be treated as a purely unresolved component or as a smooth bulge-like structure. We compare a S\'ersic+PSF model with a two-S\'ersic model that approximates a compact inner component plus an extended host. We perform these fits using \texttt{pysersic} \citep{pasha_pysersic_2023}, which enables Markov chain Monte Carlo sampling of the structural parameters of galaxies. In particular, we follow the methodology described in \citet{carreira_jwst_2026}, which describes the application of \texttt{pysersic} to imaging mosaics from JADES DR5. Here, we fit two multi-component surface brightness profiles to the Cosmic Ember: a two-S\'ersic model that treats the inner component as a bulge-like structure with smaller effective radius, and a S\'ersic+PSF model that accounts for an unresolved, PSF-like flux source in the center of the galaxy. A more detailed description of the two-S\'ersic modeling procedure can be found in \citet{carreira_jwst_2026}; the S\'ersic+PSF modeling is described in C. Carreira et al. (in prep.). 

These fits show that the Cosmic Ember contains a bright compact central component with additional structured residual emission. The morphology is therefore consistent with a strong central contribution to the PSF-matched colors, while the interpretation with no clear evidence for ongoing luminous AGN activity comes from combining this structure with the spectroscopic and MIRI constraints discussed in the main text.

\section{Additional MACER diagnostics}
\label{app:macer_comparison}

\begin{figure*}[!ht]
\centering
\includegraphics[width=\textwidth]{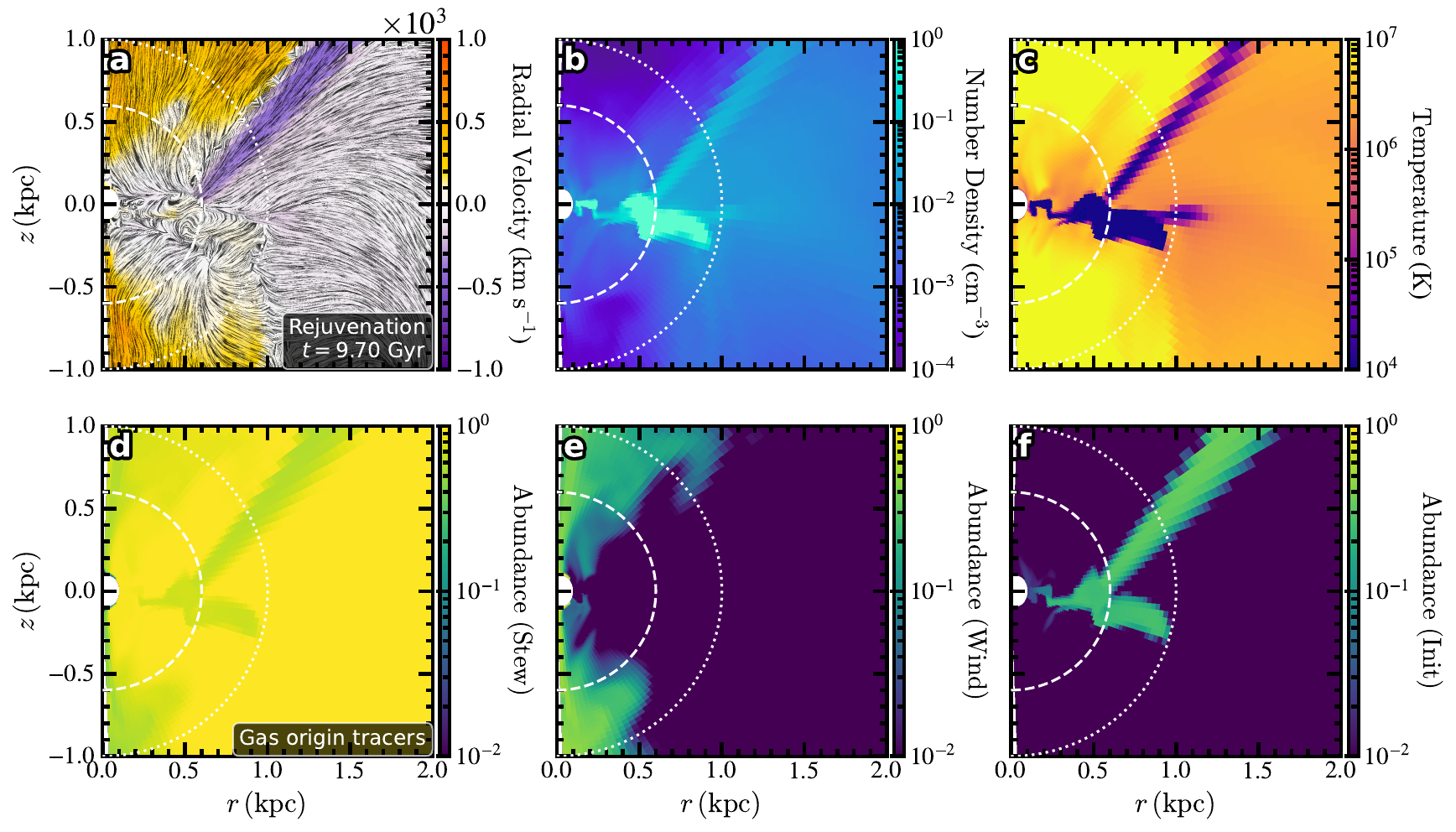}
\caption{Meridional slices at \(t=9.70\)~Gyr in the fiducial MACER run. Panels a--c show, from left to right, the radial velocity \(v_r\), gas density \(\rho\), and temperature \(T\). In panel c, the black line segments are tangential to the local velocity field and indicate the direction of the flow. Panels d--f show abundance tracers that identify the origin of the gas, corresponding respectively to stellar-wind material, AGN wind material, and the initial ISM gas. This epoch marks the central rejuvenation episode highlighted in Figure~\ref{fig:macer_ssfr}.}
\label{fig:macer_reju_slices}
\end{figure*}

\begin{figure*}[!ht]
\centering
\includegraphics[width=0.5\columnwidth]{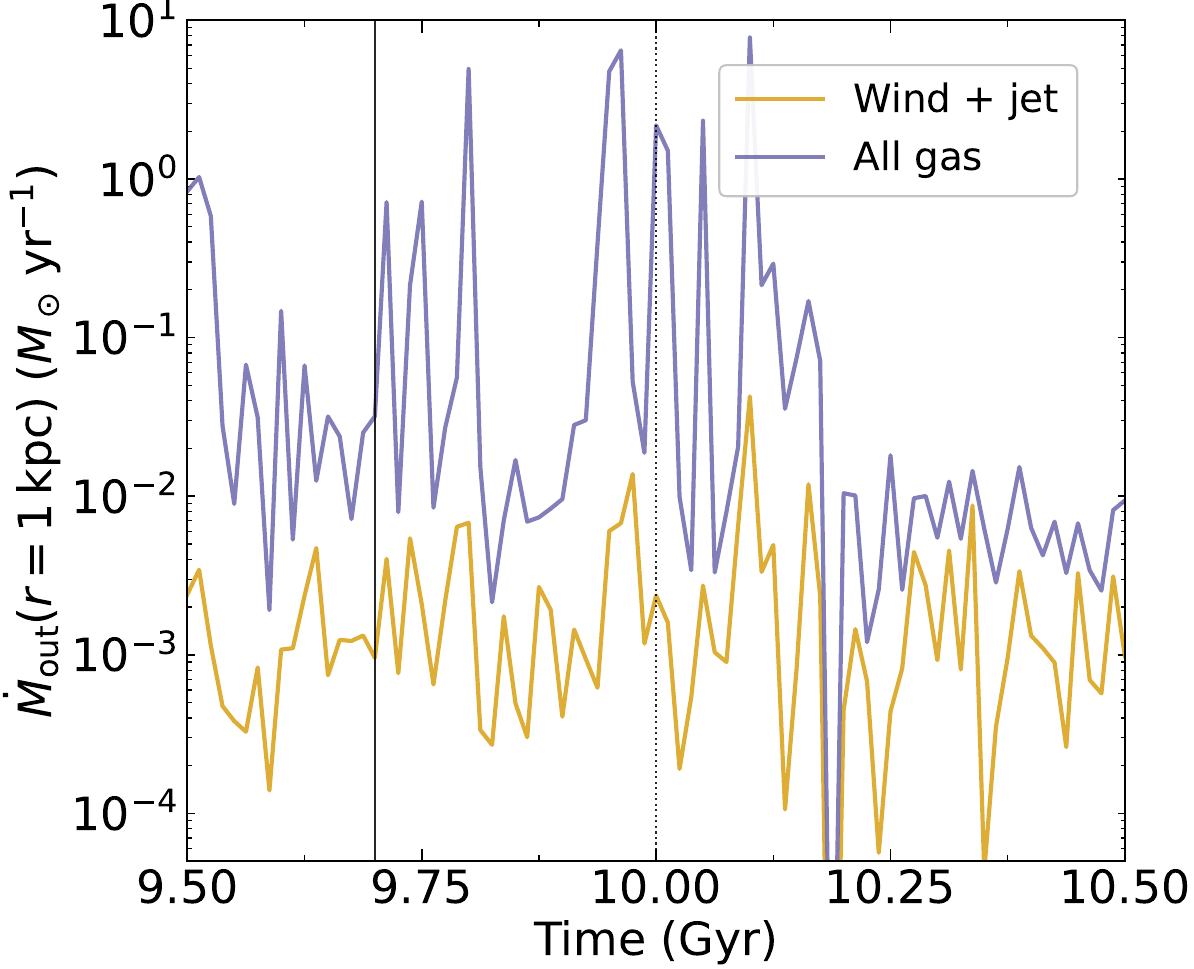}
\caption{Outward mass flux through a shell at $r=1$~kpc in the MACER fiducial run. \emph{Purple:} all gas; \emph{gold:} wind/jet-traced gas. Vertical lines mark $t=9.70$ and 10.0~Gyr, as in Figure~\ref{fig:macer_ssfr}. Near 10~Gyr the all-gas flux reaches $\dot M_{\rm out}\approx2.2\,M_\odot\,\mathrm{yr}^{-1}$, while the wind/jet contribution remains more than two orders of magnitude lower.}
\label{fig:macer_outflow_r1kpc}
\end{figure*}

\begin{figure*}[!ht]
\centering
\includegraphics[width=\textwidth]{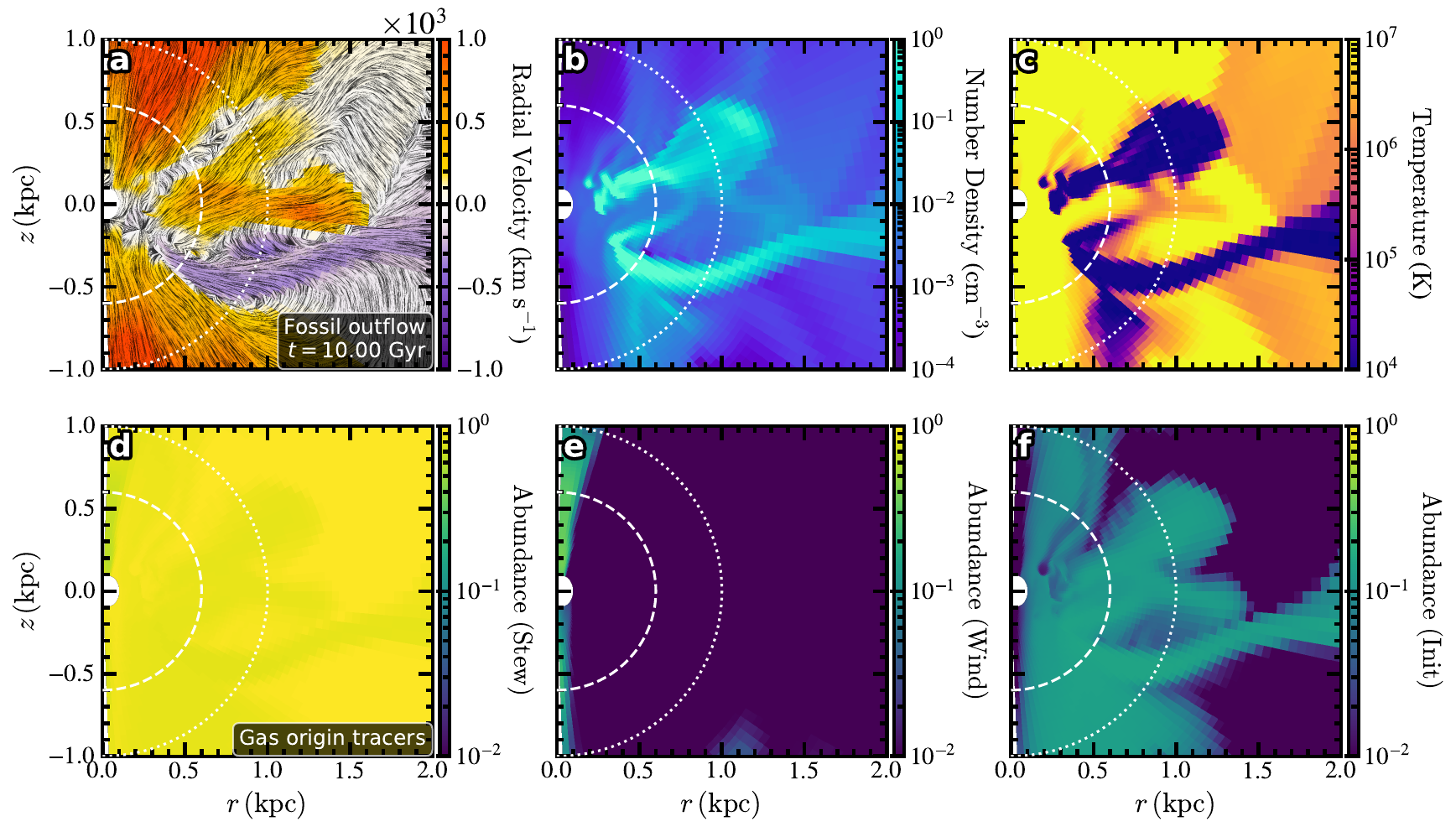}
\caption{Meridional slices at \(t=10.0\)~Gyr in the fiducial MACER run. Panels a--c show, from left to right, the radial velocity \(v_r\), gas density \(\rho\), and temperature \(T\). In panel c, the black line segments are tangential to the local velocity field and indicate the direction of the flow. Panels d--f illustrate the origin of the gas through abundance tracers, showing respectively the stellar-wind component, AGN wind material, and the initial ISM gas. Together, these panels reveal the spatial structure of the fossil wind/jet field in the inner kiloparsec.}
\label{fig:macer_fossil_slices}
\end{figure*}

Figures~\ref{fig:macer_reju_slices}--\ref{fig:macer_fossil_slices} show the MACER slice maps and mass-flux diagnostic used in Section~\ref{subsec:cosmic_ember}. These figures provide the spatial and kinematic details behind the comparison with the Cosmic Ember.

\pagebreak


\end{document}